\documentclass[reprint,amsmath,amssymb,aps,nofootinbib]{revtex4-2}

\usepackage{graphicx}% Include figure files
\usepackage{dcolumn}% Align table columns on decimal point
\usepackage{bm}% bold math
\usepackage{hyperref}% add hypertext capabilities
\graphicspath{{images/}} % images' folder
\usepackage{lipsum} % use as '\lipsum[2-4]' for dummy text
\usepackage{xcolor} % for \textcolor

\usepackage[normalem]{ulem} % \sout command to strikethrough text

\newcommand{\ve}{\vec}
\newcommand{\br}[1]{\left(#1\right)}
\newcommand{\sbr}[1]{\left[#1\right]}
\newcommand{\grad}{\nabla}
\newcommand{\ti}[1]{\tilde{#1}}

\renewcommand{\a}{\alpha}
\renewcommand{\b}{\beta}
\renewcommand{\d}{\delta}
\newcommand{\e}{\epsilon}
\newcommand{\q}{\theta} % a la Mathematica
\newcommand{\f}{\phi}
\newcommand{\m}{\mu}
\newcommand{\n}{\nu}
\renewcommand{\r}{\rho}
\newcommand{\s}{\sigma}

\newcommand{\p}{\pi}
\renewcommand{\k}{\kappa}
\newcommand{\g}{\gamma}

\newcommand{\w}{\omega}

\newcommand{\D}{\Delta}

\renewcommand{\P}{\Pi}

\begin{document}

\title{Black Hole Binaries in Cubic Horndeski Theories}

\author{Pau Figueras}%
 \email{p.figueras@qmul.ac.uk}
\affiliation{%
 School of Mathematical Sciences, Queen Mary University of London\\
 Mile End Road, London, E1 4NS, United Kingdom
}%

\author{Tiago França}
 \email{t.e.franca@qmul.ac.uk}
\affiliation{%
 School of Mathematical Sciences, Queen Mary University of London\\
 Mile End Road, London, E1 4NS, United Kingdom
}%

\date{\today}% It is always \today, today,

\begin{abstract}
We study black hole binary mergers in certain cubic Horndeski theories of gravity, treating them fully non-linearly. In the regime of validity of effective field theory, the mismatch of the gravitational wave strain between Horndeski and general relativity (coupled to a scalar field) can be as large as $10-13\%$ in the Advanced LIGO mass range. The initial data and coupling constants are chosen such the theory remains in the weakly coupled regime throughout the evolution. In all cases that we have explored, we observe that the waveform in the Horndeski theory is shifted by an amount much larger than the smallness parameter that controls the initial data. This effect is generic and may be present in other theories of gravity involving higher derivatives.
\end{abstract}

\maketitle

\section{\label{sec:intro}Introduction}

Detection of gravitational waves produced in stellar mass black hole binary mergers have revolutionised both the experimental and theoretical studies of gravity. On the one hand, the direct detection of gravitational waves allows for the possibility to perform new tests of general relativity (GR) in the strong field regime\footnote{In this article, by the `strong field regime' we mean the regime in which the non-linearities of the theory are important.} and, perhaps, observe deviations from the established theory. To do so, one needs to be able to compute theoretical waveforms from black hole binary mergers in alternative theories of gravity, see e.g., \cite{Johnson-McDaniel:2021yge}. This necessity has prompted a lot of activity in recent years to study dynamics of black holes in such theories. On the other hand, the current experimental data suggests that the potential deviations from GR are small. Whilst it is unlikely that the present generation of gravitational wave observatories can detect such deviations, the future third generation of detectors and/or the space based detectors will provide more data and with higher precision. Therefore, it is not impossible that even small deviations from GR can be detected in the future. However, there is a lack of full gravitational waveforms in alternative theories of gravity; this implies some deviations from GR may be undetected. 

To theoretically compute waveforms produced in the strong field regime one needs to work with a theory that has a well-posed initial value problem. This is not an issue if the corrections to GR are treated perturbatively. In fact, in recent years, significant progress has been made in studies of the strong field dynamics of certain alternative theories of interest using this approach \cite{Okounkova:2017yby,Witek:2018dmd,Okounkova:2018pql,Okounkova:2019dfo,Okounkova:2019zjf,Okounkova:2020rqw,Bezares:2021dma}. However, it is well-known that perturbation theory may break down over sufficiently long times due to secular effects.\footnote{There are recent interesting attempts to re-sum the perturbative series and hence alleviate these secular effects \cite{GalvezGhersi:2021sxs}.} In addition, perturbation theory may miss certain non-perturbative effects which, even if very small, may be detectable over sufficiently many orbits of a binary, as future generations of gravitational wave detectors expect to be able to observe. Therefore, it is of interest to treat alternative theories of gravity fully non-linearly and uncover some of their (perhaps) unique physical effects that may break certain degeneracies. 

In this article, when we discuss alternative theories of gravity, we refer to the `strongly coupled regime' of the theory as the regime in which the new terms in the equations of motion that modify GR are comparable (or even larger) to the original (two-derivative) terms. Conversely, by the `weakly coupled regime' we will mean the regime of the theory in which the modifications to the GR equations of motion are small. This is compatible with still being in the strong field regime of gravity. It is in the weakly coupled regime that alternative theories of gravity make sense as low energy effective field theories (EFTs).

Up until recently, only the so-called scalar-tensor and the scalar-vector-tensor theories of gravity had been studied fully non-linearly \cite{Healy:2011ef,Barausse:2012da,Hirschmann:2017psw,Sagunski:2017nzb}. The reason is that for this class of theories, it is straightforward to find a well-posed formulation. For other, more general, classes of theories involving higher derivatives and yet second order equations of motion, such as Horndeski or Lovelock theories, finding a suitable well-posed formulation turns out to be far more difficult. In fact, it has been shown that weak hyperbolicity can fail in Lovelock \cite{Reall:2014pwa} or Horndeski \cite{Ripley:2019aqj,Ripley:2019hxt,Bernard:2019fjb,Ripley:2019irj,Ripley:2020vpk,Bezares:2020wkn,Figueras:2020dzx} theories if the spacetime curvature and/or the derivatives of the scalar field become too large, i.e., in the strongly coupled regime. In a recent breakthrough, \cite{Kovacs:2020pns,Kovacs:2020ywu} showed that these theories can be strongly hyperbolic in certain modified generalised harmonic coordinates in the weakly coupled regime, i.e., when the deviations from GR are small. These theoretical developments have led to the first studies of the fully non-linear dynamics of black holes in a particular subset of these theories, namely scalar Einstein-Gauss-Bonnet theory \cite{East:2020hgw,East:2021bqk}. 

A more general approach to find well-posed formulations of general alternative theories of gravity has been proposed by \cite{Cayuso:2017iqc,Allwright:2018rut}. This proposal is inspired by the M\"uller-Israel-Stewart (MIS) formulation of relativistic viscous hydrodynamics \cite{Muller:1967aa,Israel:1976213,Israel:1976tn,Israel:1979wp}, and in principle can work even for theories with higher-than-second order equations of motion, as recently shown in \cite{Cayuso:2020lca} for a certain eight-derivative theory of gravity and in \cite{Bezares:2021yek,Lara:2021piy} for scalar tensor theories of gravity, albeit with two derivative equations of motion. Therefore, given these recent theoretical developments, it is the right time to start probing the non-linear regime of alternative theories of gravity and infer predictions for black hole binary mergers. 

Building on our previous work \cite{Figueras:2020dzx}, in this paper we study black hole binary mergers in cubic Horndeski theories. Horndeski theories are the most general scalar-tensor theories of gravity with second order equations of motion arising from a diffeomorphism invariant action in four spacetime dimensions \cite{Horndeski:1974wa,Nicolis:2008in}. Letting $\phi$ be a real scalar field, then the general action for the Horndeski theories is given by
\begin{equation}\label{eq:Horndeski_general}
    \mathcal{S} = \frac{1}{16\pi G}\int dx^4 \sqrt{-g} \br{\mathcal{L}_1+\mathcal{L}_2+\mathcal{L}_3+\mathcal{L}_4+\mathcal{L}_5},
\end{equation}
with
    \begin{align}
        \mathcal{L}_1 =&~ R + X - V(\f)\,,\nonumber\\
        \mathcal{L}_2 =&~ G_2(\f,X)\,,\nonumber\\
        \mathcal{L}_3 =&~ G_3(\f,X)\,\square\f\,,\nonumber\\
        \mathcal{L}_4 =&~ G_4(\f,X)\,R \\
        &+ \partial_X G_4(\f,X)\big[\br{\square\f}^2
        -\br{\grad_\m\grad_\n\f}\br{\grad^\m\grad^\n\f}\big]\,,\nonumber\\
        \mathcal{L}_5 =&~ G_5(\f,X)G_{\m\n}\grad^\m\grad^\n\f \nonumber\\
        & - \frac{1}{6}\partial_X G_5(\f,X)\,\big[\br{\square\f}^2  -3\square\f\br{\grad_\m\grad_\n\f}\br{\grad^\m\grad^\n\f} \nonumber\\
        &\hspace{2.5cm}+ 2\br{\grad_\m\grad_\n\f}\br{\grad^\n\grad^\r\f}\br{\grad_\r\grad^\m\f}\big]\,,\nonumber
    \end{align}
where $R$ and $G_{\mu\nu}$ are the Ricci scalar and the Einstein tensor respectively constructed from the spacetime metric $g_{\mu\nu}$, $X:= -\frac{1}{2}(\partial \phi)^2$ and $V(\phi)$ is an arbitrary potential for the scalar field, which may include a mass term. Here $G_i(\phi,X)$, $i=2,\ldots,5$ are arbitrary functions of their arguments. In this notation, $\mathcal{L}_1$ is the standard Lagrangian corresponding to GR coupled to a scalar field with potential $V(\phi)$, and $\mathcal{L}_i$, $i=2,\ldots,5$, can be interpreted as higher derivative corrections to GR in the matter sector, in this case comprised solely by a scalar field $\phi$. The cubic Horndeski theories are given by setting $G_4=G_5=0$ in \eqref{eq:Horndeski_general}. The reason why we consider these theories in the present paper and in our previous work \cite{Figueras:2020dzx} is because these theories are known to be well-posed in the standard gauges used in numerical GR \cite{Kovacs:2019jqj}. As we previously mentioned, the general case has also been shown be well-posed in \cite{Kovacs:2020pns,Kovacs:2020ywu}, but in a modified version of the generalised harmonic coordinates. While here we only consider the cubic case for simplicity and convenience, one may expect that some of our conclusions hold for more general Horndeski theories. 

In our previous paper \cite{Figueras:2020dzx} we studied gravitational collapse of a massless scalar in spherical symmetry in certain cubic Horndeski theories\footnote{Whilst in this paper the initial data was chosen to be spherically symmetric, our code did not assume spherical symmetry.} given by the choices
\begin{align}
        G_2(\f,X) &= g_2\, X^2, \label{eq:choiceG2}\\
        G_3(\f,X) &= g_3\, X, \label{eq:choiceG3}
\end{align}
where $g_2$ and $g_3$ are arbitrary dimensionful coupling constants that we can tune. This particular choice of $G_2$ is well-motivated by EFT \cite{Weinberg:2008hq}, while this choice of $G_3$ is a matter of simplicity and convenience. Both choices have been extensively considered in the literature (e.g. \cite{Brahma:2020eqd}). One of the main results of \cite{Figueras:2020dzx} was to identify the region in the space of initial conditions and couplings such that the solution in the domain of dependence of the initial data surface remained in the weakly coupled regime of the theory on and outside black hole horizons if any are present, see Section \ref{sec:methods} for more details. This is relevant in the context of EFT to justify that one can consistently keep only the leading order terms beyond GR, i.e., Horndeski, and neglect the otherwise (presumably) infinite number of higher derivative corrections. At the same time, it is consistent to treat the theory fully non-linearly, as we do here. 

In the present paper we consider the same theories as in \cite{Figueras:2020dzx}, namely \eqref{eq:choiceG2}--\eqref{eq:choiceG3}. For the initial data, we choose two boosted lumps of scalar field, with amplitudes chosen so that they quickly collapse into black holes, thus forming a black hole binary. Whilst most of the scalar field is absorbed by the black holes during the initial collapse stage, a scalar cloud remains in their vicinity throughout the lifetime of the binary. This scalar cloud can interact with itself and with the black holes and, over sufficiently long times, give rise to interesting effects. Furthermore, since spacetime curvature can source the scalar field via the Einstein equations, it is conceivable that when the spacetime curvatures are large, i.e., in the merger phase of a binary, one can observe sizeable deviations from GR. We mostly consider massive scalar fields and we restrict ourselves to a choice of potential $V(\phi)=\frac{1}{2}m^2\phi$, where $m$ is the mass of the scalar field. The reason is that the corresponding scalar cloud can remain in the vicinity of the black holes for longer and hence there is a greater chance of producing larger deviations from GR. The initial separations and velocities of the scalar lumps are tuned so that the black holes that form describe an eccentric binary that merges in 5 orbits. As we shall see in Section \ref{sec:results}, eccentric binaries seem to be particularly well-suited to detect small deviations from GR since the system enters the strong field regime in every close encounter of the binary and not only in the merger phase. It is not clear if circular binaries would also exhibit a build up of the deviations from GR during the inspiral phase and not only in the merger phase. If they do, then one should expect an even larger deviations from GR. We leave this interesting problem for future work.

Finally, we choose the coupling constants $g_2$ and $g_3$ such that on the initial data surface, the solution lies well inside the weakly coupled regime and we monitor that the solution remains in this regime throughout the evolution; this is necessary to ensure the consistency of the truncated EFT.

Note that for the cubic Horndeski theories, the natural frame to consider is the Einstein frame. This would not be the case had we considered more general Horndeski theories such as $\mathcal{L}_4$ \cite{Faraoni:1999hp}. On top of this, since massive scalar fields cannot propagate to the wave-zone, the waveforms presented in Section \ref{sec:results} would look the same in the Einstein and Jordan frames respectively. 

The rest of the paper is organised as follows: In Section \ref{sec:methods} we describe our methods, numerical techniques and construction of suitable initial data. Section \ref{sec:results} contains the main results of the paper. In Section \ref{sec:waves}, we present the waveforms computed in various cubic Horndeski theories and we compare them to the waveforms obtained in GR coupled to a scalar field. In Section \ref{sec:scalar_cloud} we discuss the properties of the scalar cloud surrounding the black holes and in Section \ref{sec:wcc} we show that in our simulations, the weak coupling conditions are satisfied throughout the evolution of the binaries. In Section \ref{sec:mismatch_results} we analyse the mismatch between GR and Horndeski. In Section \ref{sec:conclusions} we summarise the main results of the paper and we discuss future directions for research. The convergence tests are presented in Appendix \ref{appendix:convergence}.

We adopt the following notation: Greek letters ($\m, \n, \r, \dots$) to denote full spacetime indices and Latin letters ($i, j, k, \dots$) for purely spatial indices. We adopt the mostly plus metric signature and $G=c=1$. Mass scales with respect to the ADM mass of the spacetime.

\section{\label{sec:methods}Methods}
\subsection{Equations of motion}

The equations of motion for the theories that we consider in this article are given by equations $(2.2)-(2.3)$ in our previous paper \cite{Figueras:2020dzx}. In our numerical implementation, we write these equations in the usual 3+1 conformal decomposition form and use the CCZ4 formulation of the Einstein equations \cite{Alic:2011gg,Alic:2013xsa} (see also \cite{Bernuzzi:2009ex}). We use the replacement $\k_1\to\k_1/\a$ and fix the constraint damping parameters to $\k_1=0.1,\k_2=0,\k_3=1$. The matter contribution to the Einstein equations as well as the scalar field equations of motion written in the 3+1 form can be found in Appendix A of \cite{Figueras:2020dzx}. 

\subsection{Initial Data}
For initial data, we consider the superposition of two boosted equal scalar field bubbles. Each individual scalar bubble is spherically symmetric and has some in-going momentum, which prompts a quick collapse into a black hole without any outgoing scalar wave while leaving some leftover dynamical scalar hair surrounding the black hole. Whilst the individual scalar bubbles satisfy the Hamiltonian and momentum constraints, the superposition of the two does not; however, we place them sufficiently far apart so that the initial constraint violations due superposing the two scalar bubbles are sufficiently small. Further details and explicit form of the constraint satisfying scalar bubble profiles used can be found in \cite{Figueras:2020dzx}.

For binary systems, we boost the individual profiles with a Galilean boost with velocity $\ve{v}$. This can be implemented in the scalar momentum by adding to it (with the appropriate sign) the Galilean boost:
\begin{equation}
    \P(t,\ve{x})\big|_{t=0} = \P_{\text{original}}(t,\ve{x})\big|_{t=0} -\tfrac{1}{\alpha}\ve{v}\cdot\ve{\grad}\phi(t,\ve{x})\big|_{t=0}\,,
\end{equation}
where $\a$ is the lapse function. Such a boost is valid for small velocities and avoids the obstacle of having to evaluate the initial data at different times for distinct points, as one would need to do to implement a proper Lorentz boost. Doing the latter would be unfeasible for non-static initial data.

To superpose the initial data of two scalar bubbles, $A$ and $B$, boosted in opposite directions, we use:
\begin{equation}
    \begin{aligned}
        \psi &= \psi_A + \psi_B - 1\,,\\
        K_{ij} &= \g_{m(i}\br{K_{j)n}^A\g_A^{nm} + K_{j)n}^B\g_B^{nm}}\,,\\
        \phi &= \phi_A + \phi_B\,,\\
        \Pi &= \Pi_A + \Pi_B\,,
    \end{aligned}
    \label{eq:init_dat}
\end{equation}
where $\psi$ is the conformal factor associated with the induced metric on the initial data surface, $\g_{ij}$, so that $\g_{ij} = \psi^4\ti{\g}_{ij}$ and $\tilde \gamma_{ij}$ is flat. The initial data corresponding to each of the individual scalar bubbles is conformally flat, so $\tilde \gamma_{ij}=\tilde\gamma^A_{ij}=\tilde\gamma^B_{ij}=\delta_{ij}$, and satisfies $K^A=K^B=0$. For $K=0$ and a conformally flat spatial metric, the condition for $K_{ij}$ reduces to $\ti{A}_{ij} = \ti{A}_{ij}^A + \ti{A}_{ij}^B$, where $\ti{A}_{ij}$ is the conformal traceless part of the extrinsic curvature given by $\tilde A_{ij}=\psi^{-4}K_{ij}$. In our simulations, we initialise the lapse and shift vector as $\a = 1$ and $\b^i = 0$.

Motivated by the similarity with previous studies of binary black holes mergers in scalar field environments, we also tried an alternative way of constructing initial data by superposing scalar bubbles as suggested in \cite{Helfer:2021brt}. The construction of \cite{Helfer:2021brt} produces initial data which is physically different from \eqref{eq:init_dat} but, for the large separations between the initial scalar bubbles as in our paper, the amounts of initial constraint violations are comparable. Therefore, henceforth we will only discuss the evolution of the binaries constructed using \eqref{eq:init_dat}.

Using the notation of \cite{Figueras:2020dzx}, the initial scalar bubbles have parameters $(A,r_0,\w) = (0.21, 5, \sqrt{\tfrac{1}{2}})$ and a scalar mass parameter $m=0.5$. For an isolated scalar bubble, this configuration has an ADM mass of approximately $M\approx 0.53$. The mass term in the scalar potential accounts for about $10\%$ of the total ADM mass, while the contribution of the Horndeski terms for a coupling of $g_2=0.02$ or $g_3=0.05$ is of order $\mathcal{O}(10^{-5})$.

The binaries with the eccentric orbits presented in this paper are obtained by choosing a ``large" initial separation\footnote{This ``large" initial separation makes a circular binary unfeasible with our computational resources, but it helps to minimise the errors from the initial data superposition.} between the centres of the scalar bubbles of $D=40$ and an individual scalar boost velocity of $|\ve{v}| = 0.17$; after the initial gravitational collapse, the resulting black holes have an initial velocity of $0.042$. We calculate numerically that the superposed data has a total ADM mass of $M=1.0346\pm0.0001$. The values quoted above are in code units, and the mass parameter $m$ and couplings $g_2$, $g_3$ will be referred without units henceforth (e.g., $g_2=0.02$ as opposed to $g_2\sim0.0214\,M^{-2}$ after taking into account that the total mass is $M=1.0346$).

\subsection{Weak coupling conditions}
In order for the Horndeski theories to be valid EFTs, the basic requirement is that corrections to the two-derivative GR terms in the equations of motion on and outside black hole horizons (if there are any in the spacetime) should be small at all times. Inside black holes both GR and the Horndeski theories break down but classically this region is inaccessible to external observers. Therefore, in practice, we only monitor the weak coupling conditions (WCCs) on and outside black hole horizons.\footnote{While the weak cosmic censorship conjecture \cite{Penrose:1969pc,Christodoulou:1999ve} remains unproven in the astrophysical settings considered in this paper, we will assume that in holds. We do not find any evidence against it in our setting.} 

For the cases considered in this paper, the WCCs translate into the requirement \cite{Figueras:2020dzx}:
\begin{equation}
     |g_2 \, L^{-2}| \ll 1\,, \quad\quad |g_3\, L^{-2}| \ll 1\,,\label{eq:WFC_2}
\end{equation}
where $L$ is the typical length scale estimate for the system. This can be computed as
\begin{equation}
    L^{-1}=\text{max}\{\left|R_{\a\b\m\n}\right|^{\tfrac{1}{2}},\left|\grad_\m\f\right|,\left|\grad_\m\grad_\n\f\right|^{\tfrac{1}{2}},m\}\,,
\end{equation} 
where $R_{\a\b\m\n}$ is the spacetime Riemann tensor and $m$ is the mass parameter in the scalar field potential.\footnote{It turns out that for our choice of couplings and scalar mass parameter $m$, near the black holes the contribution of $m$ to $L$ is always smaller than the metric and scalar curvature terms.} We consider values for the couplings $g_2$ and $g_3$ based on the valid values of $\eta_2$ and $\eta_3$ displayed in Fig. 1 and 10 of our previous paper \cite{Figueras:2020dzx}. In practice, for massive scalar fields, the scalar clouds that form near black holes tend to be more extended, have lower densities and smaller gradients than in the massless case, thus allowing for larger couplings without violating the WCCs that may lead to a loss of the hyperbolicity of the scalar equations.

\subsection{Excision}
To control the gradients in the interior of black holes and avoid issues such that a potential loss of hyperbolicity of the scalar equations there, we implement the same form of excision as in \cite{Figueras:2020dzx}. More precisely, inside black holes we modify the equations of motion by smoothly switching off the Horndeski terms (see Appendix C of \cite{Figueras:2020dzx} for the details of our implementation). This procedure seems justified since the EFT is no longer valid in these regions of the spacetime. We only implement this form of excision during the initial stages of the evolution, while $t \lesssim 40M$, namely during gravitational collapse and gauge re-adjustment phases. Once the black hole has stabilised, the matter density at its centre is very small and no loss of hyperbolicity of the scalar equations occurs. Note that as long as the weak coupling conditions hold, all horizons should be close to the usual metric horizon and therefore excising a small enough region well-inside the metric horizon should not affect the physics in the domain of outer communications. 

This procedure of excision is not necessary with the resolutions that we used in this paper if the coupling is sufficiently small. However, for arbitrarily high resolutions, the scalar field gradients near the centre of the black holes would grow unbounded and one would expect the code to break down. 

\subsection{Gravitational wave extraction}
We extract gravitational waves at finite radii by projecting the Weyl scalar $\Psi_4$ onto the spin-weighted spherical harmonics on multiple spheres of fixed coordinate radius in the standard way, obtaining the multipoles $\psi_{\ell m}$. Data on each integration sphere is obtained from the finest available level in the numerical grid at a given extraction radius, using fourth-order Lagrange interpolation. We calculate the Weyl scalar $\Psi_4$ using the Newman-Penrose formalism \cite{Baker:2001sf} and the electric and magnetic parts of the Weyl tensor, $E_{ij}$ and $B_{ij}$ \cite{alcubierre}. The latter can be computed from our evolution variables using the following expressions adapted to the 3+1 CCZ4 formulation of the Einstein equations:
\begin{align}
    E_{ij} &= \big[R_{ij} + D_{(i}\Theta_{j)} + (K-\Theta)K_{ij} - K_{im}K^{m}_{\phantom{m}j} \nonumber\\
    &\hspace{.5cm}- \tfrac{\kappa}{4}S_{ij}\big]^{\text{TF}}\,, \label{eq:Eij}\\
    B_{ij} &= \epsilon_{mn(i}D^m K_{j)}^{\phantom{j)}n}\,,\label{eq:Bij}
\end{align}
where $R_{ij}$ is the 3 dimensional Ricci tensor, $K_{ij}$ the extrinsic curvature, $\Theta:=-n_\m Z^\m$ is the projection of the CCZ4 vector onto the timelike unit normal vector $n^\m$, $S_{ij}:=\g_i^{~\m}\g_j^{~\n}T_{\m\n}$ the spatial projection of the stress-energy tensor, $\g_{\m\n} := g_{\m\n}+n_\m n_\n$ is the induced metric on the spatial hypersurfaces, $\e_{\m\n\r} = n^\s \e_{\s\m\n\r}$ is the volume form on such hypersurfaces and $[\cdot]^{TF}$ denotes the trace-free part of the expression in square brackets. Note that equations \eqref{eq:Eij}-\eqref{eq:Bij} guarantee $E_{ij}$ and $B_{ij}$ are automatically trace-free and symmetric, unlike usual 3+1 ADM expressions \cite{alcubierre}, which require that the constraints are satisfied.

\subsection{Gravitational strain}
The natural observable measured in detectors and used when constructing templates is the gravitational strain. The strain of a gravitational wave, $h$, can be obtained from the $\Psi_4$ Weyl scalar using the transformation \cite{Radia:2021smk}:
\begin{equation}
    \Psi_4 = -\ddot{h} = -\ddot{h}^+ + i \ddot{h}^\times\,,
\end{equation}
where the dot $\dot{\,}$ denotes a time derivative, and $h^+$ and $h^\times$ are the usual plus and cross polarisations of the wave. This gives the strain multipoles $\ddot{h}^+_{\ell m} = -\text{Re}(\psi_{\ell m})$ and $\ddot{h}^\times_{\ell m} = \text{Im}(\psi_{\ell m})$, where $\psi_{\ell m}$ are the amplitudes of each mode in the multipolar decomposition of the Weyl scalar $\Psi_4$. To avoid artefacts from finite length of the wave, discrete sampling and noisy data, we perform the double time integration in the frequency domain using a fixed frequency filter \cite{Reisswig:2010di}, with a cutoff of $0.01M^{-1}$ for low frequencies.\footnote{This choice affects the noise in the strain, but low frequencies have negligible effects in the final computation of the mismatch. Furthermore, adding a cutoff for the high frequencies resulted in no improvement.} We taper the signal in the time domain with a Tukey window \cite{McKechan:2010kp} of width $\sim 40M$ on each side\footnote{This choice reduces noise, but it does not affect the results in any meaningful way, as the signal is essentially zero in this region.} and zero-pad to the nearest power of two. We further zero-pad the waveform to increase the length by a factor of eight before applying the fast Fourier transform \cite{Varma:2018mmi}, in order to increase the frequency resolution of the discrete Fourier transform and reduce the noise for low frequencies introduced by the discretization. Removing the initial junk radiation of the inspiral from the time domain did not result in any significant improvement. Finally, the signal at null infinity can be obtained by extrapolating the results at finite radii assuming a Taylor series expansion in $\frac{1}{r^*}$ \cite{Sperhake:2010tu}, where $r^*=r + 2M \log\left|\frac{r}{2M}-1\right|$ is the tortoise radius, after first aligning separate extraction radii in retarded time, $u = t - r^*$.

\subsection{Waveform mismatch}\label{sec:mismatch_theory}
In order to estimate the difference between two waveforms, we compute the mismatch between the strain resulting from each wave. First, given the strain of two waves, $h_1(t)$ and $h_2(t)$, one can compute the overlap, $\mathrm{O}$, using the frequency domain inner product \cite{Blackman:2017dfb, Owen:1995tm, Lindblom:2008cm}:
\begin{align}
    \mathrm{O}(h_1, h_2) &= \frac{\text{Re}\br{\left<h_1,h_2\right>}}{\sqrt{\left<h_1,h_1\right>\left<h_2,h_2\right>}}\,,\\
    \left<h_1,h_2\right> &= 4\int_{f_{\text{min}}}^{f_{\text{max}}} \frac{\ti{h}_1^*(f)\ti{h}_2(f)}{S_n(f)}df\,,
\end{align}
where $\ti{h}(f)$ denotes the Fourier transform of the function $h(t)$, $^*$ denotes complex conjugation, $S_n(f)$ is the power spectral density (PSD) of a detector's strain noise as a function of frequency $f$ (e.g., updated Advanced LIGO sensitivity design curve \cite{aLIGOupdatedantsens}), $f_{\text{min}}$ and $f_{\text{max}}$ is the lowest and highest frequency cutoffs of the PSD of the detector or the frequency minimum/maximum imposed by the timestep and duration of the simulation. Notice that for $h_1=h_2$, $\left<h_1,h_2\right>$ is real.

Then, we compute the mismatch by maximising the overlap over time and phase shifts of the second wave, $h_2^{\d t,\f}(t) = h_2(t+\d t)e^{i\f}$:
\begin{align}
    \text{mismatch} &= 1 - \underset{\d t,\f}{\text{max}}~\mathrm{O}(h_1, h_2^{\d t,\f})\,.
\end{align}
Noticing that $\ti{h}_2^{\d t,\f}(f) = \ti{h}_2(f) e^{i\f}e^{2\p i\d t}$ and
\begin{align}
    \mathrm{O}(h_1, h_2^{\d t, \f}) = \frac{\text{Re}\br{e^{i\f}e^{2\p i\d t}\left<h_1,h_2\right>}}{\sqrt{\left<h_1,h_1\right>\left<h_2,h_2\right>}}\,,
\end{align}
then maximising over phase shifts corresponds to simply taking the absolute value of $\left<h_1,h_2\right>$ as opposed to the real part. Maximising over time shifts is more subtle because the discrete domain implies the Fourier transform changes by more than a mere phase $e^{2\p i\d t}$. Hence, we perform the time shift maximisation numerically. To allow for continuous time shifts, we interpolate the data $h_{1,2}(t)$ and re-sample appropriately after the time shift. The number of points when re-sampling the time series does not affect the final result. When comparing two gravitational waves, the length of the time interval used for the Fourier transform and the size of the frequency domains used for integration are the same for all waves (taking into account time shifts). All in all:
\begin{align}
    \text{mismatch} &= 1 - \frac{\underset{\d t}{\text{max}} ~\left|\left<h_1,h_2^{\d t}\right>\right|}{\sqrt{\left<h_1,h_1\right>\left<h_2,h_2\right>}}\,.
\end{align}

\subsection{Numerical scheme}
The equations of motion are evolved with \texttt{GRChombo} \cite{Clough:2015sqa, Andrade2021, Radia:2021smk}, using MPI, OpenMP and templated SIMD/vector intrinsics to obtain a good performance in the most common architectures. \texttt{GRChombo} uses the \texttt{Chombo} adaptive mesh refinement libraries \cite{chombo}. We use a tagging criterion that triggers the regridding based on second derivatives of both the scalar field and the conformal factor. \texttt{GRChombo} implements the usual puncture gauge for the evolution of the gauge variables and $N=3$ \textit{Kreiss-Oliger} (KO) dissipation, with fixed $\s=1$ in all our simulations. As of boundary conditions, we use Sommerfeld boundary conditions and take advantage of the reflective/bitant symmetry of the binary problem to evolve only half of the grid. Sixth order spatial stencils are used in order to improve phase accuracy of the binaries \cite{Husa:2007hp}. Time updates are still made with a fourth-order Runge-Kutta scheme, which implies that the global convergence order cannot be higher than four.\footnote{Notice that this allows us to use the usual KO dissipation stencils that are commonly implemented with fourth order finite differences.} For the results presented in this paper, we have a \textit{Courant-Friedrichs-Lewy} factor of $1/4$, a coarsest level resolution of $\D x = \frac{16}{7}$, with 8 additional refinement levels, and a computational domain of size $1024^3$.

\section{\label{sec:results}Results}
In this section we present the results of our numerical simulations for the various Horndeski theories that we have considered and we compare them to GR. To carry out the comparison, we consider standard GR coupled to a massive scalar field (with same mass parameter $m=0.5$ as in Horndeski). We comment on the massless scalar field case in Section \ref{sec:scalar_cloud}. 
We have also considered the evolution of a black hole binary in vacuum GR with the same total ADM mass and initial velocities for the black holes. In this case, the binary describes many more orbits before merger, as expected, since no energy is transferred to the scalar field. We will not comment any further on this case since it is not relevant for the kind of comparisons that we carry out.

We have constructed superposed initial data for GR coupled to a massive scalar field and for Horndeski theories. One could question whether different results arise from small differences in the initial data. As discussed in \cite{Figueras:2020dzx}, the effect of the Horndeski terms in the initial data is proportional to $\frac{g_2A^2}{r_0^2}$ and $\frac{g_3Aw^2}{r_0^4}$ depending on the theory. For the values of the couplings $g_2$ and $g_3$ that we consider, this results in a difference of order $\mathcal{O}(10^{-5})$ between the Horndeski and the GR counterpart. To confirm that the small Horndeski corrections in the initial data do not affect the subsequent evolution, we evolved the equations of motion of the Horndeski theories using initial data constructed for GR. Clearly this procedure introduces extra initial constraint violations proportional to the Horndeski couplings. However, our results from the Horndeski theories initialised with GR initial data and those results obtained using proper Horndeski initial data do not exhibit any significant or quantitative difference. Therefore, we conclude that the differences observed between GR and Horndeski theories are caused by the evolution with distinct evolution equations and not by the extremely small differences in the initial data. Henceforth, for the Horndeski theories we will only present results obtained with Horndeski initial data.
\begin{figure}[t]
\centering
\includegraphics[width=0.45\textwidth]{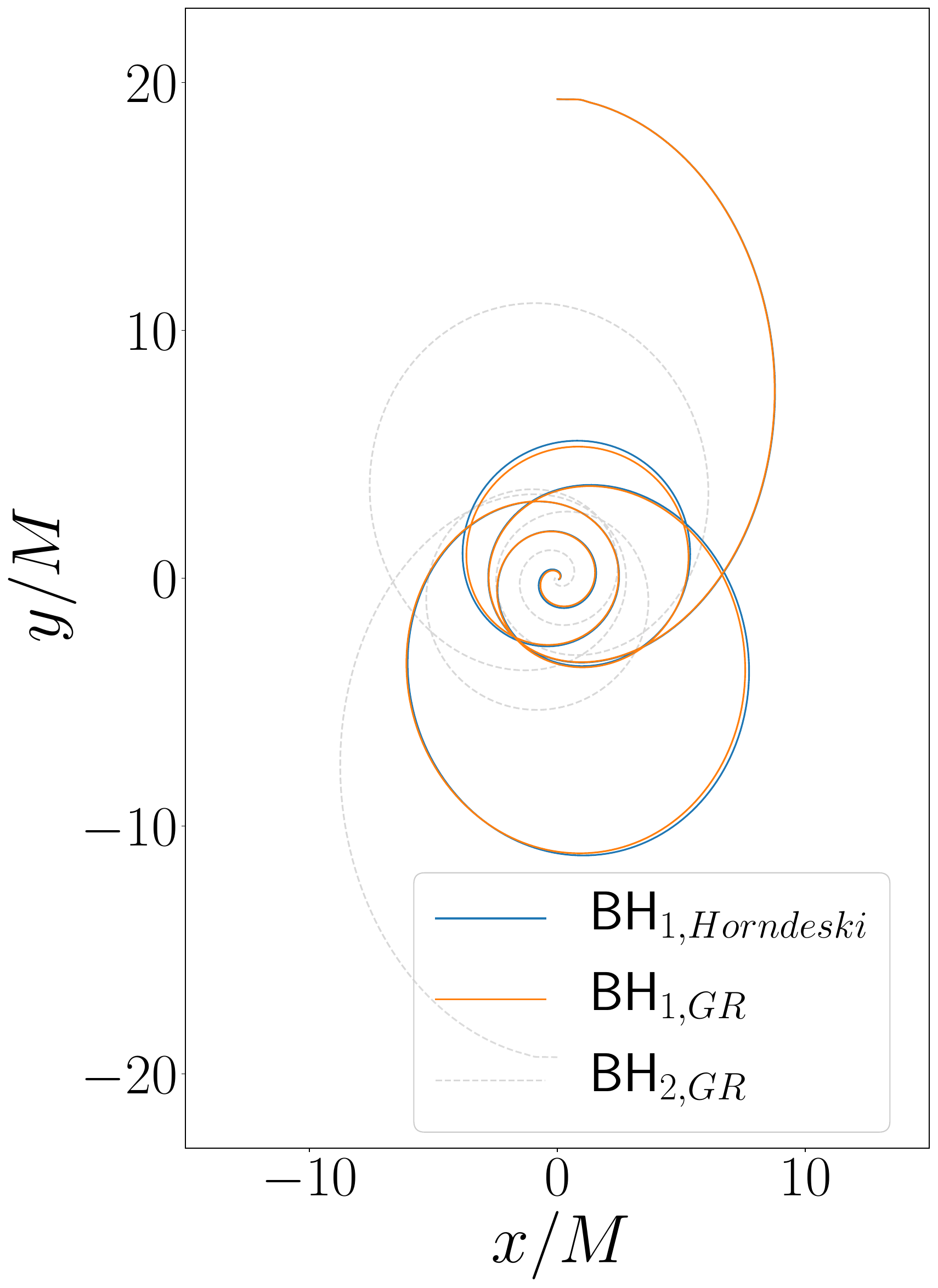}
\caption{Orbits of the two black holes in Horndeski for $g_2=0.02$ and GR. For clarity, in the Horndeski case we only show one of the black holes.}
\label{fig:AH_xy}
\end{figure}

In Fig. \ref{fig:AH_xy} we display the trajectories of the punctures on the orbital plane for GR and for a Horndeski theory with $g_2=0.02$. This figure shows that after the first close encounter of the binary, the trajectories that the black holes follow in GR and in Horndeski are visibly different. Interestingly, the black holes seem to recombine to the same trajectory in the final stages of binary. In the following subsections we will quantify the differences in other observables such as the gravitational strain.

\subsection{Waveform strain}\label{sec:waves}
\begin{figure*}
\centering
\includegraphics[width=0.98\textwidth]{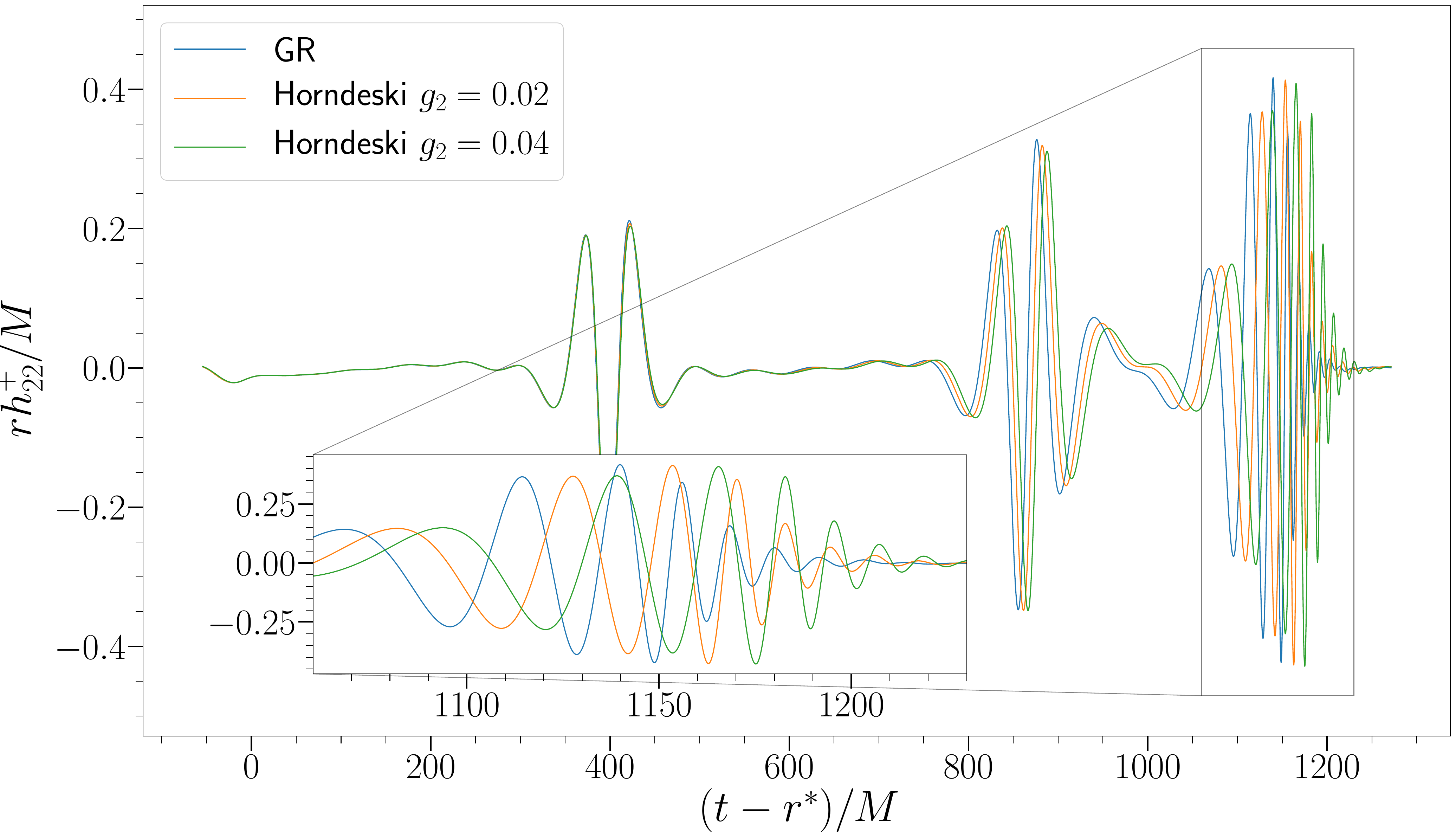}
\caption{\label{fig:g2_0.02_gw}Comparison of gravitational wave between Horndeski theory with $g_2=0.02$, $g_2=0.04$ and GR in retarded time, $u=t-r^*$, where $r^*$ is the tortoise radius. Displaying the $(\ell,m)=(2,2)$ mode of the plus polarisation of the strain, $h^+_{22}$, extrapolated to null infinity. There is a visible misalignment between GR and Horndeski that builds up over time, becoming larger during the merger phase.}
\end{figure*}

\begin{figure*}
\centering
\includegraphics[width=0.98\textwidth]{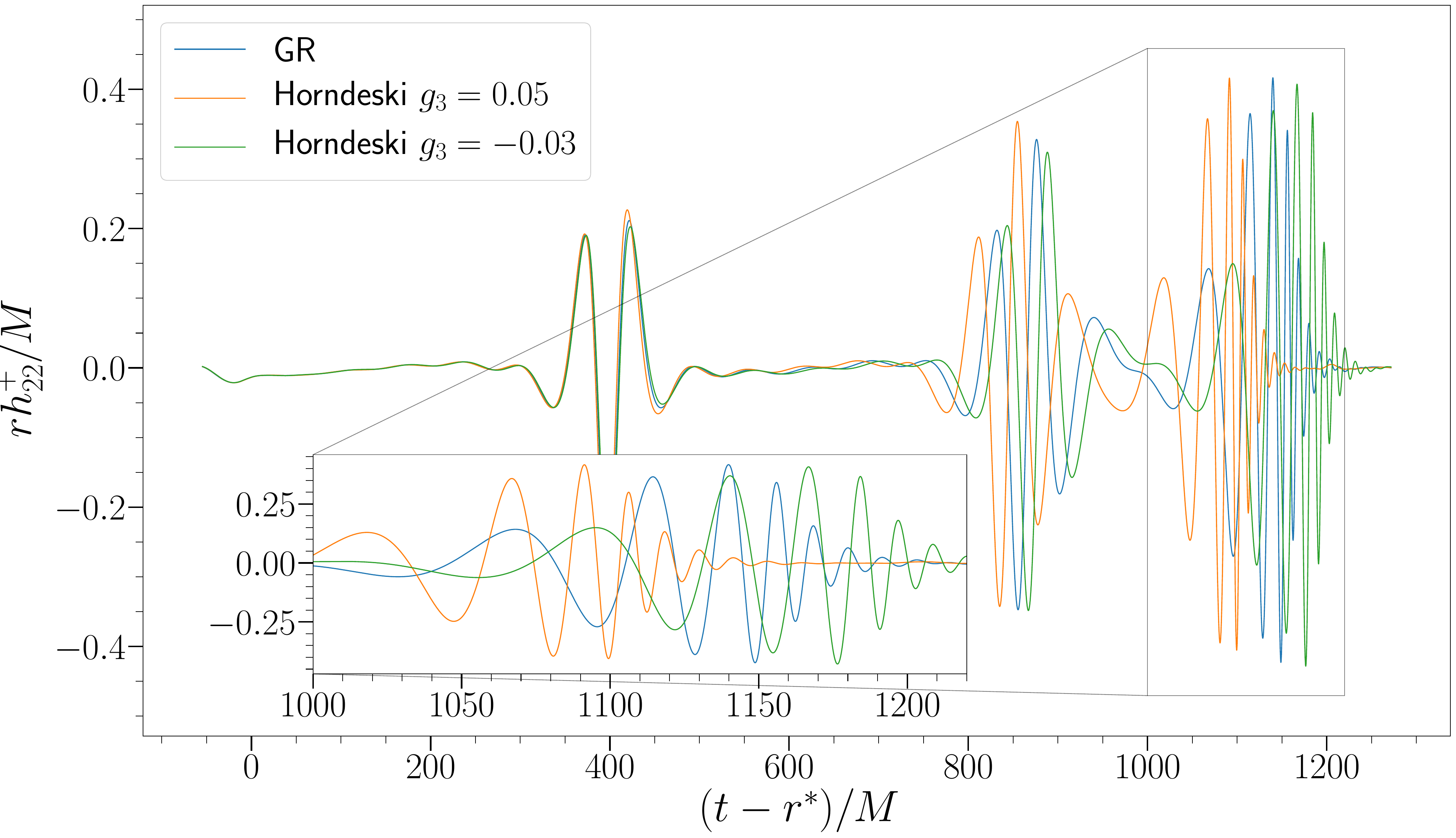}
\caption{\label{fig:g3_gw}Comparison of gravitational wave between GR and Horndeski theories with $g_3=0.05$ and $g_3=-0.03$, in retarded time, $u=t-r^*$, with $r^*$ the tortoise radius. Displaying the $(\ell,m)=(2,2)$ mode of the plus polarisation of the strain, $h^+_{22}$, extrapolated to null infinity. As in the $G_2$ theory, Fig. \ref{fig:g2_0.02_gw}, we observe a misalignment that builds up over time.}
\end{figure*}

In this subsection we compare the waveform strain for eccentric binaries in GR and in different Horndeski theories. For the latter, we consider both the $G_2$ and the $G_3$ theories for different values and signs of the coupling constants. In Figs. \ref{fig:g2_0.02_gw} and \ref{fig:g3_gw} we present the $(\ell,m)=(2,2)$ mode of the plus polarisation of the strain, $h^+$, extrapolated to null infinity using 6 radii between $50-150M$, for the $G_2$ and $G_3$ theories respectively. The strain for higher $(\ell,m)$ modes exhibits qualitatively similar features.

Referring to Figs. \ref{fig:g2_0.02_gw} and \ref{fig:g3_gw}, the two peaks that can be seen at $t\sim 400M$ and $t\sim 850M$ correspond to the bursts of radiation emitted during the first two close encounters of the eccentric binary\footnote{It may be useful for the reader to match the gravitational wave signal in these figures with the visual animation of one of our simulations: \href{https://www.youtube.com/watch?v=uOed4AG1ulg}{https://www.youtube.com/watch?v=uOed4AG1ulg}.} before the final merger phase. The latter starts at around $t\sim 1100M$ and ends by $t\sim 1200M$, depending on the theory and the value and sign of the coupling constants. As for the final state, since the class of theories that we consider do not admit hairy black holes \cite{Hui:2012qt, Maselli:2015yva}, the end state of the evolution is a Kerr black hole surrounded by a scalar cloud. For the runs shown in Figs. \ref{fig:g2_0.02_gw} and \ref{fig:g3_gw}, the estimated parameters of the final black holes are summarised in Table \ref{tab:runs}. Note that any junk radiation caused by the initial constraint violations or choice of initial data is very small on the scale of these figures, but still visible in the first $\sim 50M$.
\begin{table}[h]
\caption{\label{tab:runs}%
Parameters of the final state Kerr black hole for each coupling $g_2$ and $g_3$. Values estimated from the final apparent horizon and errors estimated from the differences between medium and high resolutions.}
\begin{ruledtabular}
\begin{tabular}{cccccc}
\textrm{Coupling}&
\textrm{Final Mass $M_F/M$}&
\textrm{Spin Parameter $a/M$}\\
\colrule
$\text{GR}$ & $0.973\pm0.001$ & $0.676\pm0.001$ \\
$g_2=~~0.005$ & $0.973\pm0.001$ & $0.676\pm0.001$ \\
$g_2=~~0.02~$  & $0.973\pm0.001$ & $0.675\pm0.001$ \\
$g_2=~~0.04~$  & $0.973\pm0.001$ & $0.673\pm0.001$ \\
$g_3=~~0.05~$  & $0.975\pm0.001$ & $0.680\pm0.001$ \\
$g_3=-0.03~$  & $0.972\pm0.001$ & $0.672\pm0.001$
\end{tabular}
\end{ruledtabular}
\end{table}

As Figs. \ref{fig:g2_0.02_gw} and \ref{fig:g3_gw} show, the waveforms obtained in GR and in the various Horndeski theories that we considered, coincide during the initial stages of the binary, but a clear misalignment builds up over time, starting from the second close encounter of the binary and becoming more pronounced in the merger phase. This misalignment is much larger than the smallness parameter controlling the weak coupling conditions of the initial data. In subsection \ref{sec:wcc} we will provide evidence showing that a suitable local weak coupling condition remains small during the whole evolution of the binary and hence, in our setting, the Horndeski theories should be valid (and predictive) classical EFTs. 
The large misalignment that we observe in Figs. \ref{fig:g2_0.02_gw} and \ref{fig:g3_gw} is a cumulative effect arising from the locally small differences between GR and Horndeski, and it gets enhanced whenever the system enters the strong field regime, which happens in each close encounter of the eccentric binary and in the merger phase. This is expected since the corrections to GR are sourced by spacetime and scalar curvature and those become more important precisely in the strong field regime. 
Therefore, eccentric binaries seem to be useful to potentially detect deviations from GR sourced by curvature through the built up of small cumulative effects and their enhancement in the close encounters. It is conceivable that linearising the Horndeski theories around GR may allow one to compute some of the misalignment (at least for some small enough couplings) during the merger phase since its duration is relatively short and secular effects may not be an issue. However, it seems unlikely that such an approach would be able to capture the cumulative large deviations that arise from successive close encounters of an eccentric binary, such as in the examples considered here. The relatively long times that we have evolved the binaries require a full non-perturbative treatment of the theory to avoid potential secular effects. 

For the $G_2$ theory \eqref{eq:choiceG2}, a positive $g_2$ coupling induces a delay of the waveform when compared to GR, whilst a negative $g_2$ gives rise to an advancement of the signal. On the other hand, for the $G_3$ theory \eqref{eq:choiceG3} the effect is the opposite: a positive $g_3$ coupling leads to an advancement of the signal while a negative $g_3$ leads to a delay when compare to the GR waveform. In general, the observed misalignment between GR and Horndeski seems to be a rather generic effect that does not depend on the specifics of the theory. Of course, the details such as the amount or the sign of the deviations will depend on the details theory under consideration. Therefore, we are tempted to conjecture that gravitational strain computed in general Horndeski theories of gravity that do not admit equilibrium hairy black holes but with dynamical long-lived scalar clouds surrounding black holes will be misaligned with respect of the GR signals. Finally we note that the peak amplitude of the waveforms seems to be very similar across all theories and couplings. We will point out in the Section \ref{sec:conclusions} how this misalignment may be potentially detected in gravitational wave observations.

Note that the final state of GR and Horndeski simulations for $g_2\le0.04$ seems to have the same exact mass and only tiny differences in spin (see Table \ref{tab:runs}). This is counter-intuitive given the differences that the gravitational waveforms exhibit and it could be related to the fact that the trajectories of the black holes in the two theories visibly differ in the intermediate stages of the binary, but coincide again near the merger phase (see Fig. \ref{fig:AH_xy}). The physical mechanisms behind this observation may be related to the frequency shifts analysed in subsection \ref{sec:mismatch_results}. The fact that the initial and final state coincide and yet the waveforms are different indicates that, at least for equal mass non-spinning binaries, the degeneracy between the class of Horndeski theories that we have considered and GR is broken. This suggests that the degeneracy between GR and Horndeski may also be broken for unequal mass non-spinning configurations. It would be interesting to study the effects of the intrinsic spins in alternative theories of gravity. 
\begin{figure}
\centering
\includegraphics[width=0.48\textwidth]{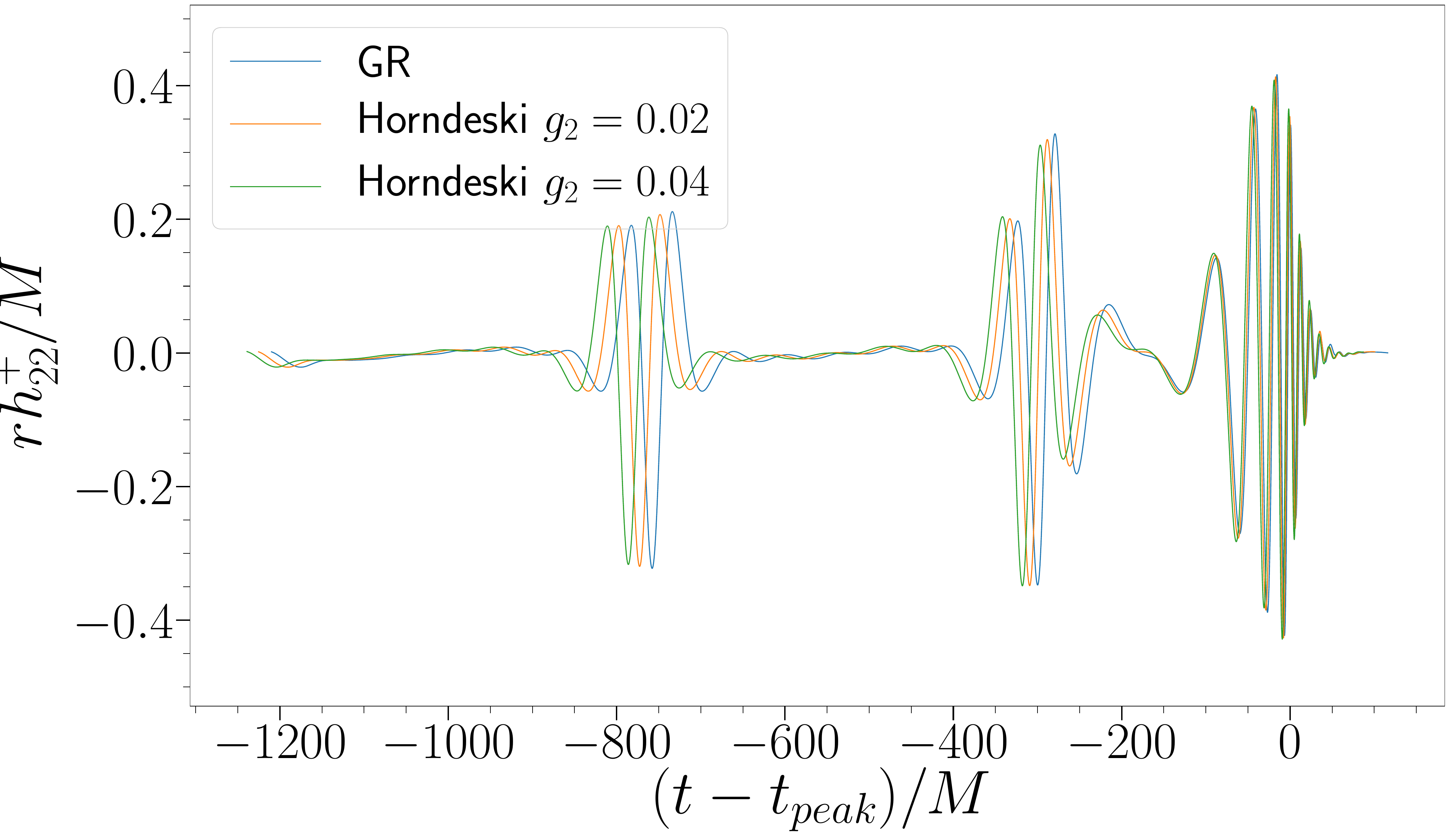}
\caption{\label{fig:g2_0.02_gw_aligned}Comparison of re-aligned $h^+_{22}$ between Horndeski theory with $g_2=0.02$, $g_2=0.04$ and GR. The waves were aligned so the peak of the amplitude of the complex strain coincides.}
\end{figure}

When comparing the waveforms between different theories, one might alternatively want to align the main peaks. However, clearly the misalignment would not disappear; it would simply be translated along the time axis. This can be seen in Fig. \ref{fig:g2_0.02_gw_aligned}, where the misalignment is now seen at the early encounters of the inspiral. This shows that the gradual phase shift is a physical effect that cannot be ignored by a constant phase shift and does not depend on how one does the comparison. For long lived inspirals beyond the strong field regime simulated with numerical relativity, the effect would be enhanced and the misalignment would be present regardless of the time or phase shift considered.

\subsection{\label{sec:scalar_cloud} Scalar Cloud}
\begin{figure}
\centering
\includegraphics[width=0.48\textwidth]{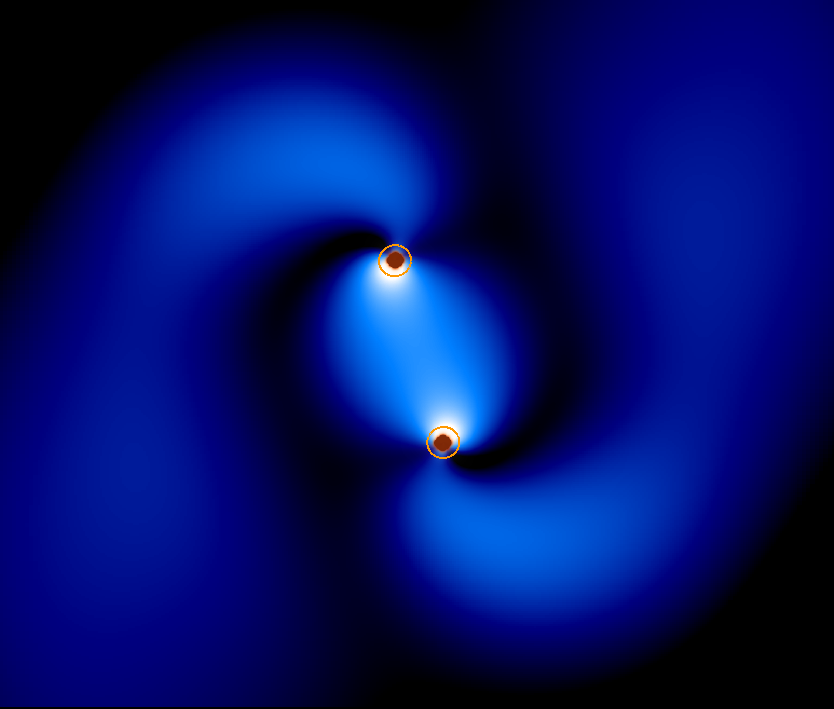}
\caption{Energy density (in blue) of the scalar field surrounding the binary black holes for the Horndeski theory with $g_2=0.005$ at a representative instant of time during the inspiral phase. The apparent horizon of the black holes is shown in orange. The region where the weak coupling conditions are larger than one is depicted in brown. Clearly this region is contained well inside the apparent horizon, as required.}
\label{fig:results_eta2}
\end{figure}

In Fig. \ref{fig:results_eta2} we display a snapshot of a binary for the $G_2$ theory with $g_2=0.005$ at a representative instant of time before the merger. This figure shows that the energy density of the scalar field (in blue) is localised in the region near the black holes, being largest near the horizons. It is in this regions where the spacetime and scalar field curvatures are largest, even though the WCCs remain small on and outside the black holes. 

For the Horndeski theories that we considered, the accumulation of non-linear effects is possible due to the presence of long lived scalar cloud surrounding the black holes. This scalar cloud survives all the way up to and well beyond the merger. This is due to the presence of a mass term in the scalar potential, since it is well-known (see e.g., \cite{Press:1972zz,Barranco:2012qs,Barranco:2017aes,Hui:2019aqm,Clough:2018exo,Ikeda:2020xvt}) that the effective potential that the scalar field ``sees" has a wall that makes it difficult for it to escape to infinity. A scalar mass parameter of $m=0.5$ is comparable to what has been seen to give rise to long lived scalar clouds \cite{Ikeda:2020xvt}, though the effects observed in this paper did not require any fine tuning. We have also carried out simulations of binaries with massless scalars and the absence of a significant scalar cloud trivially removes any long term effects of the scalar field on the evolution. 
\begin{figure}[t]
\centering
\includegraphics[width=0.48\textwidth]{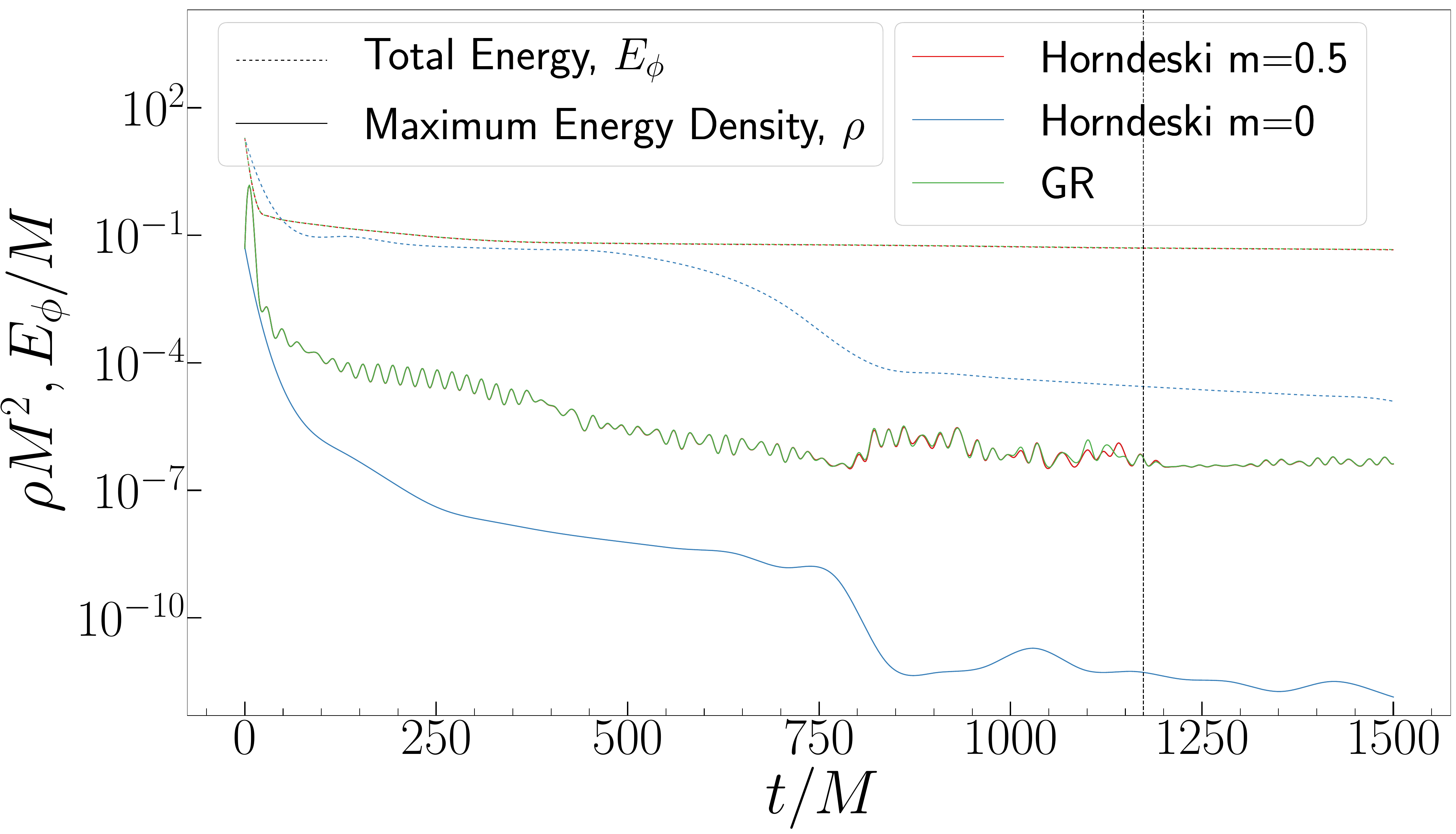}
\caption{Total energy of the scalar field $E_\phi$ and maximum value of energy density $\rho$ on the spacetime (excluding black holes, by removing from the volume of integration the interior of apparent horizons) for Horndeski with $g_2=0.005$ and a massive scalar field (in red), the same theory with a massless field (in blue) and GR with a massive scalar field (in green). A dashed black line is used to indicate the estimated merger time for the Horndeski run with a massive scalar field.}
\label{fig:energy_density_g2}
\end{figure}

In Fig. \ref{fig:energy_density_g2} we display the evolution of the total energy of the scalar field $E_\phi$ and the evolution of the maximum of the energy density $\rho$ for GR (green), for the $G_2$ theory with $g_2=0.005$ (red) and for the same Horndeski theory but with a massless scalar field (blue). After the initial gravitational collapse, most of the scalar field is absorbed by the black holes, but in the massive scalar cases, a long lived scalar cloud forms around the black holes. After the first close encounter of the eccentric binary $(t\sim 400 M)$, the maximum energy density of the scalar cloud is of order $10^{-5}M^{-2}$ for the massive scalar cases (GR and Horndeski), and it decreases very slowly with time. This long lived cloud makes it possible for the scalar field to interact with itself and with the geometry and give rise to the build up of significant differences in the physical observables such as the gravitational strain. The theories that we considered do not admit stationary hairy black holes \cite{Hui:2012qt, Maselli:2015yva} so eventually the scalar field will partly escape to (timelike) infinity and partly be absorbed by the black hole, but the timescale for this to happen is very long (much longer than our simulations). 

On the other hand, in the massless case, Fig. \ref{fig:energy_density_g2} shows that a much larger amount of scalar field is absorbed by black holes during the collapse phase. Furthermore, both the total energy of the scalar field and its energy density show a pronounced dip at $t\sim 850M$, namely in the second close encounter of the binary, indicating that any leftover amount of scalar field in the vicinity of the black holes gets absorbed. Beyond this point, the energy density of the scalar field is less than $10^{-10}M^{-2}$ while the total energy is of the order of $10^{-5}M$ (corresponding to scalar waves radiated to infinity), and both continue to steadily decrease with time. By the time the merger takes place the maximum energy density of the scalar field is compatible with numerical error. Therefore, we conclude that in the massless case, after the second close encounter of the binary, there is basically no significant amount of scalar field left in the neighbourhood of the black holes to give rise to any sizeable effect, at least for the duration of our simulations. As a consequence, no noticeable differences between GR and Horndeski are observed in the massless scalar field case.

Comparing Horndeski with GR in Fig. \ref{fig:energy_density_g2} shows that local differences (in time) in the energy density between GR and Horndeski for the massive scalar field are not significant for most of the binary, including the two close encounters; only during the merger phase one can see some differences of order $10^{-5}M^{-2}$. These results are expected if the weak coupling conditions are satisfied. Furthermore, the fact that the energy density of the massive scalar field around the black holes is small during the highly dynamical stages of the binary is necessary but not sufficient to ensure that the WCCs are satisfied.

\subsection{Weak coupling conditions during the evolution}\label{sec:wcc}
The results reported in subsection \ref{sec:waves} can only be trusted as long as the Horndeski theories that we consider are valid (truncated) EFTs. In this subsection we provide evidence that for the initial data and couplings that we considered in the paper, the local WCCs \eqref{eq:WFC_2} are satisfied at all times, thus ensuring the predictivity of the EFTs. 
\begin{figure}[t]
\centering
\includegraphics[width=0.48\textwidth]{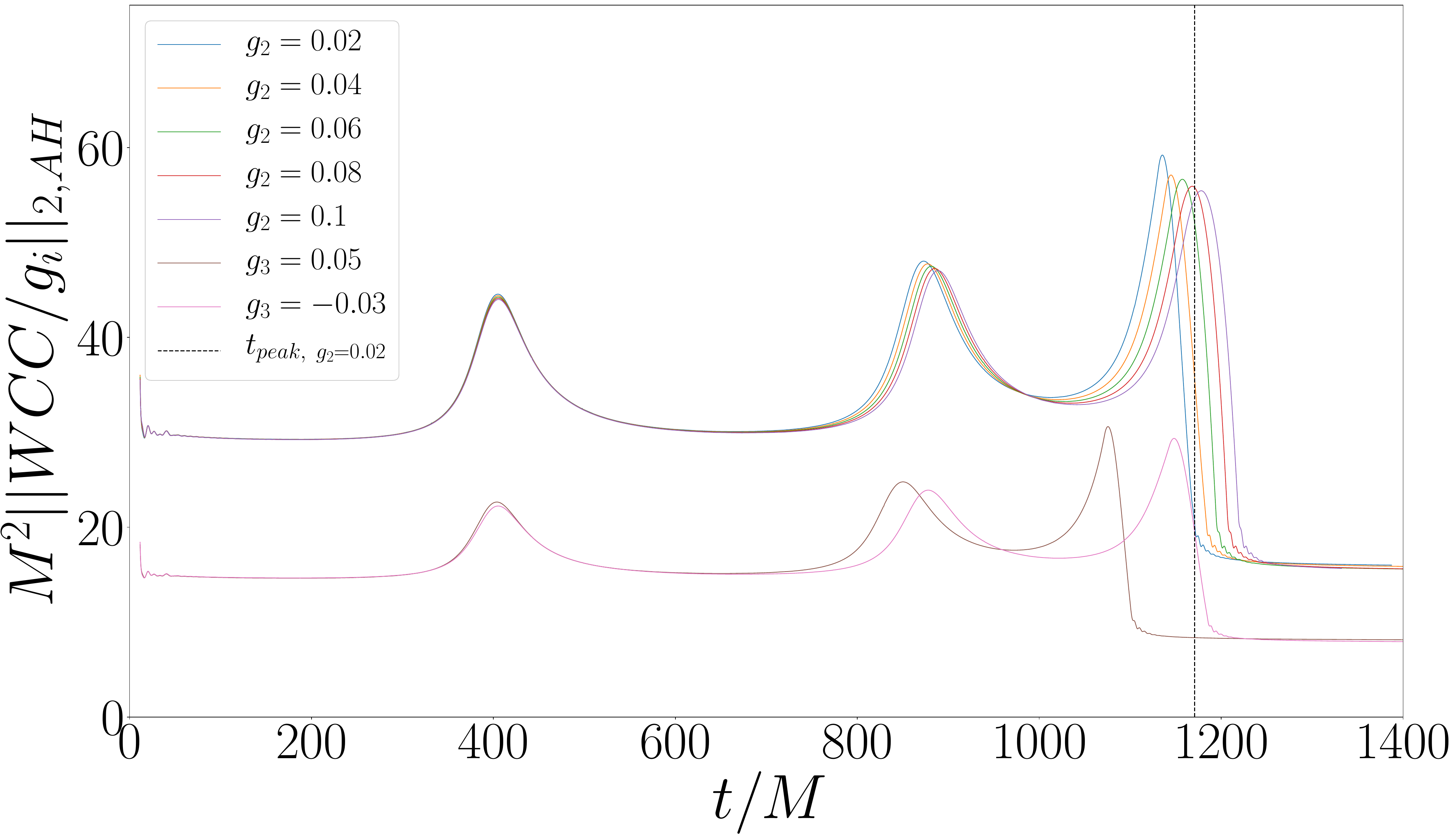}
\caption{$L^2$ norm of the weak coupling conditions \eqref{eq:WFC_2} integrated over the apparent horizon and normalised by the coupling constant $g_2$ or $g_3$. For the binary that we have evolved, this shows that $M^2 WCC/|g_2| \lesssim 50$ and $M^2 WCC/|g_3| \lesssim 20$, which in turn implies that $|g_2| \lesssim 0.02$ and $|g_3| \lesssim 0.05$ to guarantee that the WCC \eqref{eq:WFC_2} is roughly less than one. The dashed black line corresponds to the peak of the amplitude of the strain for $g_2=0.02$.}
\label{fig:WFC_spherical_L2}
\end{figure}

In Fig. \ref{fig:WFC_spherical_L2} we display the $L^2$ norm of the WCCs \eqref{eq:WFC_2} integrated on the black holes' apparent horizons, as a function of time for an eccentric binary evolved with the $G_2$ theory with different values of the coupling constant $g_2$ and one value $g_3$ coupling for the $G_3$ theory. Excluding the interior of black holes, the apparent horizons are where the WCCs have the largest values in the whole domain. This plot shows that the weak coupling condition \eqref{eq:WFC_2} remains approximately constant during the evolution, except in the close encounters of the binary and the final merger phase. The latter events correspond to the peaks in Fig. \ref{fig:WFC_spherical_L2} that can be seen at $t\sim400M$, $t\sim850M$ and $t\sim1100M$, when the system enters the strong field regime. The constancy of \eqref{eq:WFC_2} during the inspiral phase is related to the fact the energy density of the scalar field in the vicinity of the black hole remains approximately constant during this phase. The fact that the WCCs exhibits local maxima at the close encounters indicates that in an eccentric binary, we probe the strong field regime during various phases of the binary and not only near and during the merger phase as in a circular binary. It is interesting to see that when one normalises the WCCs \eqref{eq:WFC_2} by the coupling constant, the curves for the $g_2$ couplings collapse onto a single curve, except in regions where the system is in the strong field regime. This indicates that the WCC depends on the coupling constant in a trivial way when the system is not in the strong field regime.

Fig. \ref{fig:WFC_spherical_L2} shows that for our choice of initial data, $M^2 WCC/|g_2| \lesssim 50$ and $M^2 WCC/|g_3| \lesssim 20$ at all times. This implies that if we want the WCCs \eqref{eq:WFC_2} to be roughly less than one at all times, and hence guarantee that that the Horndeski theory is a valid EFT throughout the evolution, then one must choose $|g_2|\lesssim 0.02$ or $|g_3|\lesssim 0.05$. For values larger than these, the WCCs become comfortably larger than one at different (or all) stages of the binary. However, after the initial collapse stage, even for large values of the coupling $g_2$ well-beyond the regime of validity of EFT (e.g. $g_2=0.1$), the equations of motion of the scalar field remain hyperbolic throughout the inspiral and merger phases as long as the scalar density is small enough near the black holes. 

When evaluating the WCCs \eqref{eq:WFC_2} on the apparent horizon to produce Fig. \ref{fig:WFC_spherical_L2}, one has to be careful as we are actually dealing with different trapped surfaces. Due to the slicing condition used, each black hole has a trapped surface that during merger shrinks to the puncture, while a larger common apparent horizon forms, surrounding the previous ones \cite{Pook-Kolb:2019ssg,Thornburg:2006zb}. This implies that if one is computing the WCCs \eqref{eq:WFC_2} on the trapped surfaces collapsing to the punctures, it will result in unreasonably large values. To get around this gauge issue, we interpolate the data for the WCC of the original black hole apparent horizons just before merger with the data for the common apparent horizon just after it forms, excluding the unphysically large values right at the merger. The details of how one does the interpolation and which data points are excluded do not affect significantly the bounds $M^2 WCC/g_2 \lesssim 50$ and $M^2 WCC/g_3 \lesssim 20$.

We close this subsection emphasising that our assessment of the regime of validity of EFT is qualitative at best and, up to $\mathcal{O}(1)$ factors, the unity value of the WCCs is a mere order of magnitude; a more detailed study is needed in order to precisely identify this regime for the cases that we have considered. The conditions \eqref{eq:WFC_2} are only local; over time, the small effects accumulate giving rise to large deviations in some non-local observables such as the waveforms. In the context of complex scalar field with a Mexican hat type of potential, in \cite{Reall:2021ebq} is is proved that, for sufficiently long times, the truncated EFT will inevitably deviate from the UV theory. Therefore, one has to be cautious when using a truncated EFT for very long times compared to the UV mass scale, even if the local weak coupling conditions hold (see also \cite{Davis:2021oce}).

\subsection{Mismatch}\label{sec:mismatch_results}
\begin{figure}[t]
\centering
\includegraphics[width=0.48\textwidth]{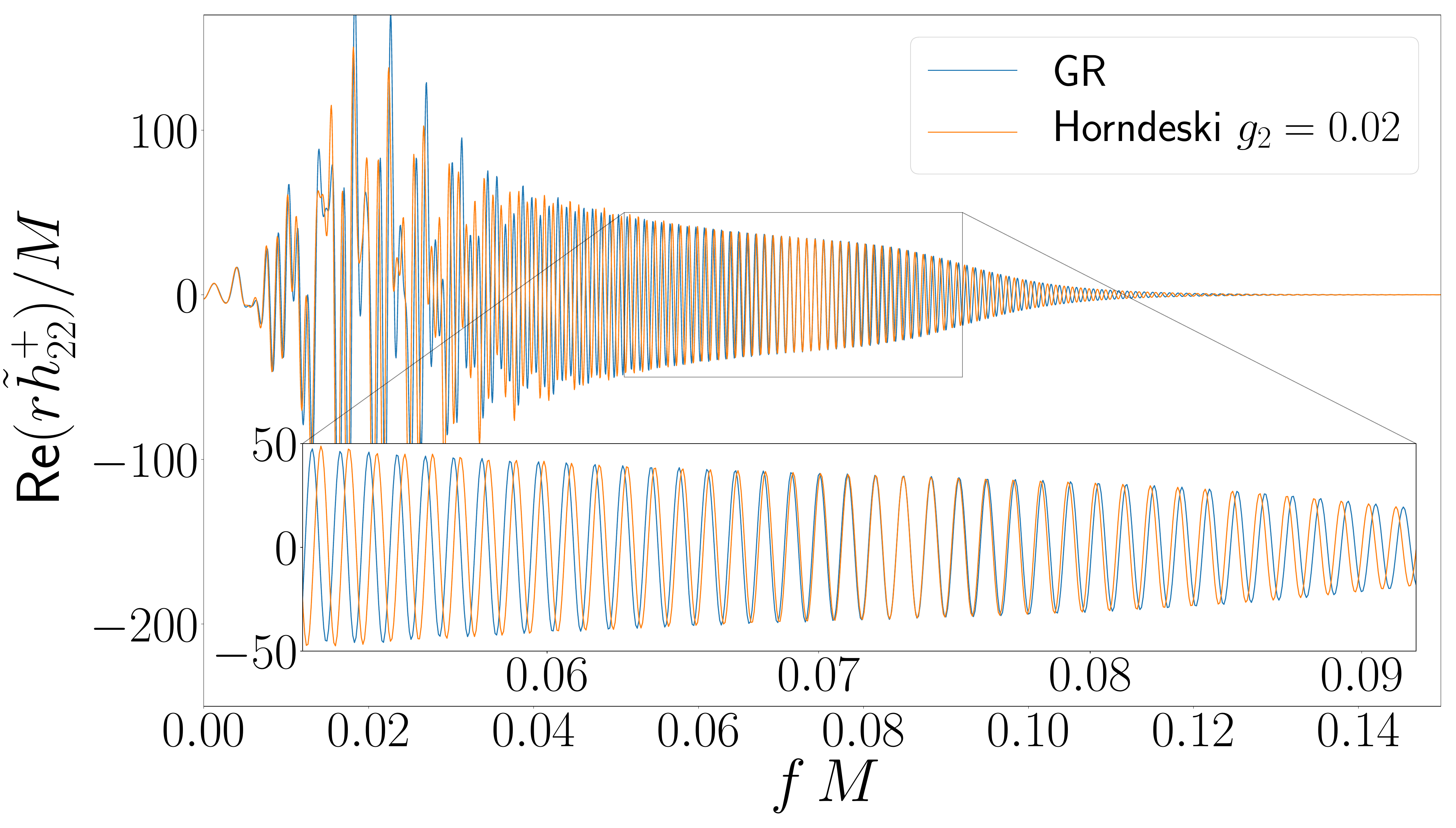}
\caption{Real part of $\ti{h}^+_{22}$, the discrete Fourier transform of $h^+_{22}$ for positive frequencies. Notice that for frequencies around $f\sim0.07M^{-1}$, Horndeski and GR align, but for lower and higher frequencies they separate in phase in opposite directions. This effect cannot be mitigated by a constant time or phase shifts of the waveform.}
\label{fig:FFT_strain_GR_vs_g2}
\end{figure}
\begin{figure}[t]
\centering
\includegraphics[width=0.48\textwidth]{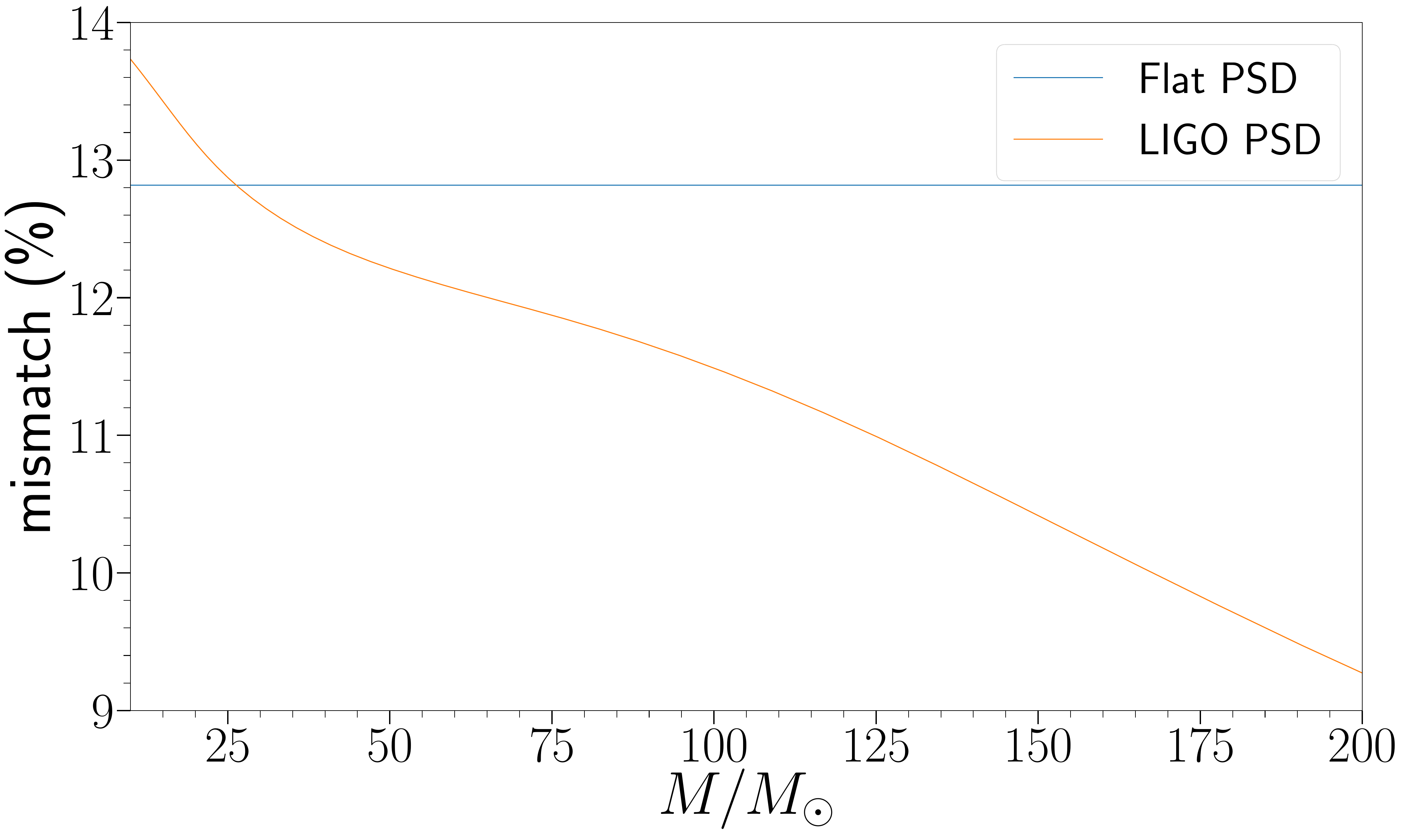}
\caption{Mismatch for $h^+_{22}$ between GR and Horndeski for $g_2=0.02$, as a function of the final black hole mass (in units of solar masses, $M_\odot$). As power spectral densities, we used the updated Advanced LIGO sensitivity design curve (\textit{aLIGODesign.txt} in \cite{aLIGOupdatedantsens}, which imposes $f_{\text{min}}=5$Hz) and a flat noise mismatches ($S_n=1$). This allows us to estimate a range for expected mismatches of $10-13 \%$.}
\label{fig:mismatches_g2_vs_GR}
\end{figure}

In this subsection we discuss our results for the mismatch between the GR and Horndeski waveforms. We focus on the $G_2$ theory with $g_2=0.02$ as an example of the limiting coupling that still satisfies the WCCs. Hence, the results for the mismatch presented should be understood as upper bounds. The mismatch depends on the coupling constants in the expected way, and the results are qualitatively the same for the $G_3$ theory.

In Fig. \ref{fig:FFT_strain_GR_vs_g2} we compare the frequencies of the real part of $\ti{h}^+_{22}$, the discrete Fourier transform of the $(\ell,m)=(2,2)$ mode of the strain, extrapolated to null infinity. Interestingly, this figure shows that in spite of both theories having approximately the same amplitudes for each frequency in the spectrum, the spectrum of the phase of the complex-valued Fourier transform differs. In range of medium frequencies, i.e., $f\sim 0.07-0.08M^{-1}$, GR and Horndeski theory agree very well. However, for both lower and higher frequencies, a significant discrepancy can be clearly seen. This effect cannot be mitigated by a constant time or phase shift of the time-domain waveform and hence we conclude that it is a physical effect. This discrepancy of both the high and low frequencies suggests that Horndeski theory exhibits both an inverse and a direct energy cascades. It would be interesting to confirm if this is indeed the case and quantify these cascades. Note that because the weak cosmic censorship holds in our scenarios, there is a natural UV cutoff for the frequencies that are accessible to external observers. As long as this cutoff is at lower energies, i.e., larger distances, than the UV cutoff of the theory, then the EFT should be valid; the fact that the WCCs hold in our case, indicates that this is indeed the case.

In Fig. \ref{fig:mismatches_g2_vs_GR} we quantify the mismatch for a detector setup receiving the plus polarisation of the strain, extrapolated to null infinity, between GR and the Horndeski theory. We restrain ourselves to using the $(\ell,m)=(2,2)$ mode, $h^+_{22}$, as this is the dominant mode an order of magnitude when compared to higher modes. We use the updated Advanced LIGO sensitivity design curve (\textit{aLIGODesign.txt} in \cite{aLIGOupdatedantsens}, which imposes $f_{\text{min}}=5$Hz) and flat noise ($S_n=1$) following the procedure described in Section \ref{sec:mismatch_theory}. We compute the mismatch for black hole masses in the typical range of stellar mass black holes binaries observed so far, $M\in[10,200]M_\odot$ \cite{LIGOScientific:2021djp}. This figure shows the mismatch varies between $\sim13\%$ at the low mass end and $\sim10\%$ at the high mass end. To confirm accuracy of these results, the mismatch between two different resolutions of the same GR evolution ranges between $0.3-0.5\%$ for the same mass ranges. As a reference, for a signal to noise ratio of 25, similar to GW150914 \cite{LIGOScientific:2016aoc}, the minimum expected mismatch for detection is about $0.6\%$ \cite{Chatziioannou:2017tdw,Purrer:2019jcp}. Additionally, in \cite{Lindblom:2008cm} is it estimated that a mismatch of $3.5\%$ would result in a $10\%$ smaller detection rate; therefore, the large mismatches obtained for big enough values of the couplings suggest that if the underlying theory of gravity was Horndeski with a massive scalar field, some events may have been undetected if the black holes had sufficient scalar field surrounding them.

\section{\label{sec:conclusions}Summary and conclusions}
In this article we have studied eccentric black hole binary mergers in certain cubic Horndeski theories \eqref{eq:choiceG2}--\eqref{eq:choiceG3} with a massive scalar field with mass parameter $m=0.5$. We have chosen initial data and small enough coupling constants such that a certain local weak coupling condition \eqref{eq:WFC_2} is satisfied at all times during the evolution. This condition monitors the size of the Horndeski terms in the equations of motion compared to GR terms, and the fact it holds ensures that the EFTs are in their regime of validity and hence we can trust their predictions. 

One of the goals of this article was to identify potential deviations from GR in some physical quantities that Horndeski theories of gravity may exhibit. We have observed that locally small deviations from GR build up over time and get enhanced whenever the system enters the strong field regime. In the case of the eccentric binaries, this happens during the successive close encounters of the black holes and in the final merger phase. Since the modifications of GR are locally small, large deviations may still arise in non-local observables, such as gravitational waveforms, through a cumulative build up. This cumulative effect gets reflected in the gravitational waveforms as large shifts with respect to the analogous waveforms computed in GR coupled to a massive scalar field with the same mass and angular momentum. 
Whilst the details, such as its sign and size, of the observed shift in the waveforms depend on the details of the theory and value of the coupling constants, the effect seems to be generic, at least within the class of Horndeski theories that we have explored here. We conjecture that the same effect should be present in more general Horndeski theories. We have quantified the misalignment of the $(\ell,m)=(2,2)$ mode of the plus polarisation of the strain, $h^+_{22}$, for one of the Horndeski theories that we have considered. We find that the spectrum differs both for low and high frequencies. Furthermore, for large enough values of the couplings, still in the regime of validity of the EFT, we find that the mismatch is around $10-13\%$ in the whole mass range of current detections. This is quite significant and it suggests that if the underlying theory of gravity differs from GR, some events where the black holes have sufficient scalar field surrounding them may have been and continue to go undetected. For smaller values of the couplings, the mismatch would be smaller.

The misalignment that we have observed is a cumulative effect and hence it only occurs if the black holes are surrounded by a long-lived scalar cloud. In our case this is possible because of the mass term in the scalar potential, which ensures that a non-trivial scalar energy density remains in the vicinity of the black holes for very long times, thus allowing the scalar field to interact with itself and with the geometry. We have also considered massless scalars, but in this case we do not observe any significant difference between Horndeski and GR. This is expected because the theories considered do not admit stationary hairy black holes and, hence, a massless scalar field gets absorbed by the black holes or disperses to null infinity on a time scale much quicker compared to the binary lifetime. In our particular example, the scalar field is essentially completely absorbed in the second close encounter of the binary and by then there has not been enough time to build up any sizeable deviation from GR. 

In this article we considered both $G_2$ and $G_3$ Horndeski theories and, as we have already mentioned, even though the initial and final states are the same, both lead to misaligned waveforms with respect to GR. Therefore, at least for equal mass non-spinning binaries, the degeneracy between the class of Horndeski theories that we have considered and GR is broken. However, we do not see any visible difference between the waveforms obtained in the $G_2$ or in the $G_3$ theories. It would be interesting to investigate if (or how) the degeneracy of the waveforms is broken in Horndeski theories of gravity. It would be interesting to extend our studies to unequal mass and spinning binaries to see if the degeneracy with GR and with the various Horndeski theories is broken when considering different mass ratios and non-zero spins. 

We have considered Horndeski theories simply as toy models for EFTs with higher derivatives; in the Horndeski case, the higher derivatives are in the matter (scalar) sector and the equations of motion are of second order. However, more fundamental theories of gravity, such as string theory, predict higher curvature corrections of the Einstein-Hilbert action. In general, such new terms in the action will result in equations of motion of order higher than two. Refs. \cite{Cayuso:2017iqc,Allwright:2018rut,Cayuso:2020lca} have outlined how the strong field regime of such theories may be probed, but it would be very interesting to do so in the context of a black hole binary. Our work suggests that in the weakly coupled regime, where these theories are valid EFTs, some of the problems that may arise in general, such as loss of hyperbolicity or shock formation, can be controlled in a physical situation that probes the strong field regimes such as a black hole binary merger. 

The main goal of the present paper is to identify what features in the physical observables extracted from black hole binaries in Horndeski theories can allow one to differentiate these theories from GR. Given that the corrections to GR have to be locally small in order for these theories be valid EFTs, non-local observables such as gravitational waveforms are particularly useful because small effects can accumulate and, for long enough times, give rise to large deviations from GR. These or other deviations from alternative theories of gravity are potentially being undetected by current gravitational wave observatories. Therefore, our results stress the importance of modelling waveforms in alternative theories of gravity treating them fully non-linearly. It would be interesting to identify other observables where large deviations show up. In the case of waveforms, until complete waveform templates are built for alternative theories, a potential way to detect the misalignment that we have identified is the following: future space-based gravitational wave observatories such as LISA are expected to be able to detect gravitational waves produced in stellar mass black hole binaries during earlier stages of the inspiral phase. From these waveforms one should be able to extract the parameters of the binary and, by using GR, predict the time of merger of the binary. Some binaries should enter the LIGO band in the final stages of the inspiral and merger phase, thus allowing to contrast the GR prediction for the merger time with the observation; a certain advancement or delay of the merger could be attributed to the fact that higher derivative corrections modify GR.

\begin{acknowledgments}
Part of the work was presented at the workshops ``Current Challenges in Gravitational Physics”, SISSA (April 2021) and ``Mathematical and Numerical Aspects of Gravitation”, IPAM, UCLA (October 2021). 
We thank the GRChombo collaboration (www.grchombo.org) for their support and code development work. We also thank Katy Clough, \'Aron Kov\'acs, Luis Lehner, Miren Radia, Josu Aurrekoetxea and Harvey Reall for useful discussions. PF is supported by the European Research Council Grant No. ERC-2014-StG 639022-NewNGR, and a Royal Society University Research Fellowship Grants No. UF140319, RGF\textbackslash EA\textbackslash180260, URF\textbackslash R\textbackslash 201026 and RF\textbackslash ERE\textbackslash 210291. TF is supported by a PhD studentship from the Royal Society RS\textbackslash PhD\textbackslash 181177. The simulations presented used PRACE resources under Grant No. 2020235545, PRACE DECI-17 resources under Grant No. 17DECI0017, the CSD3 cluster in Cambridge under projects DP128 and TC011, the MareNostrum4 cluster at the Barcelona SuperComputing Centre under Grants No. FI-2020-3-0007, FI-2020-3-0010, FI-2021-3-0010, and Athena at HPC Midlands+ cluster. The Cambridge Service for Data Driven Discovery (CSD3), partially operated by the University of Cambridge Research Computing on behalf of the STFC DiRAC HPC Facility (www.dirac.ac.uk). The DiRAC component of CSD3 is funded by BEIS capital via STFC capital grants ST/P002307/1 and ST/R002452/1 and STFC operations grant ST/R00689X/1. DiRAC is part of the National e-Infrastructure. The authors gratefully acknowledge the Gauss Centre for Supercomputing e.V. (www.gauss-centre.eu) for providing computing time on the GCS Supercomputer SuperMUC-NG at Leibniz Supercomputing Centre (www.lrz.de). The HPC Midlands cluster was funded by EPSRC on the grant EP/P020232/1, as part of the HPC Midlands+ consortium.

\end{acknowledgments}

\appendix

\section{Convergence}\label{appendix:convergence}
In this appendix we provide some details of the convergence tests that we have carried out. As a representative example, we considered the binary in the $G_2$ Horndeski theory with coupling constant $g_2=0.02$, and we performed three simulations with different resolutions to study convergence. Our simulations are evolved with a coarsest level resolution of $\D x = \frac{16}{7}$ (medium resolution), with 8 additional refinement levels and a computational domain of size $1024^3$. To carry out the tests, we used one lower resolution changing $\D x = \frac{8}{3}$ (low resolution) and one higher resolution with $\D x = 2$ (high resolution).
\begin{figure}[h]
\centering
\includegraphics[width=0.48\textwidth]{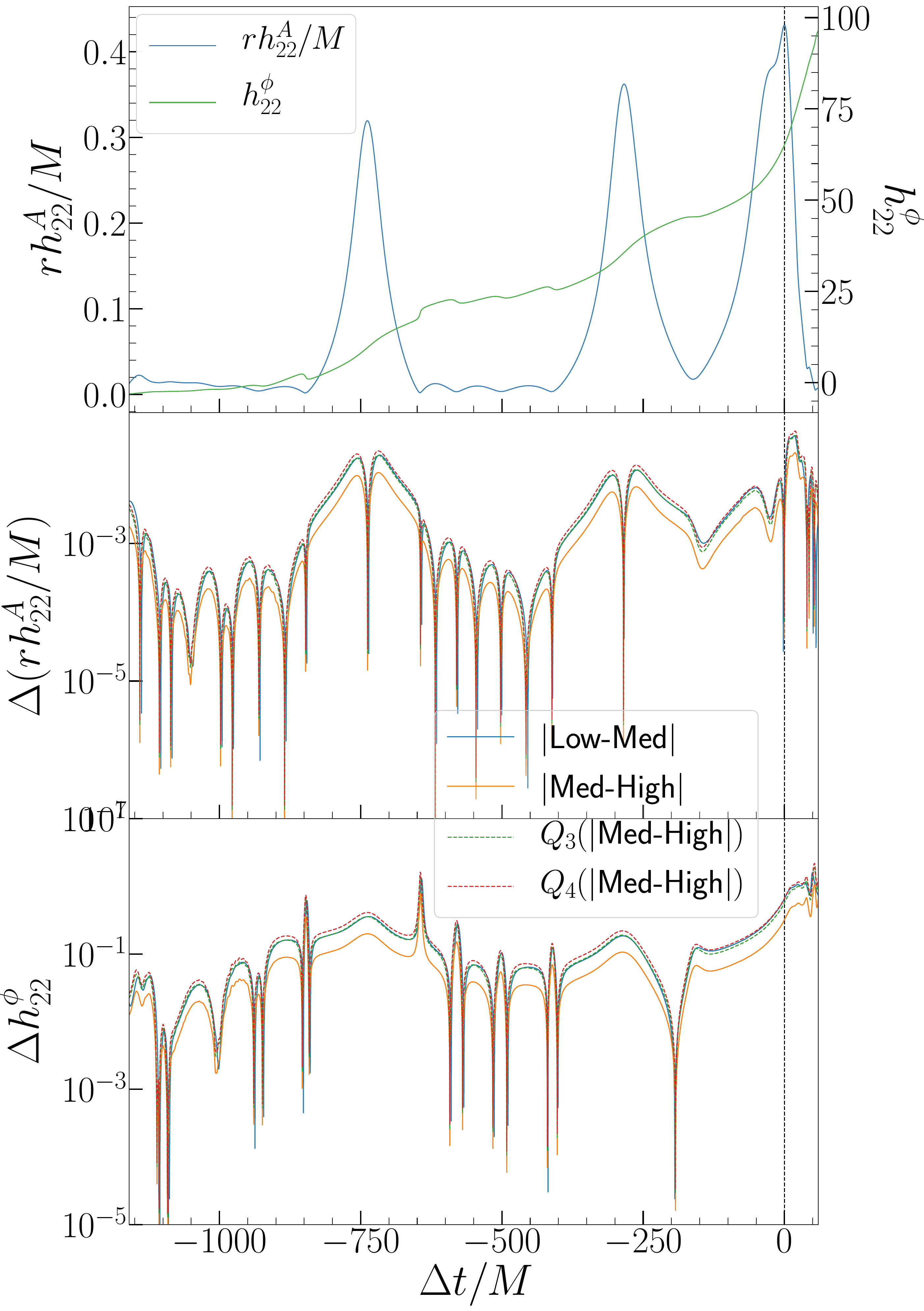}
\caption{Convergence test for $g_2=0.02$ with different coarse resolutions: low ($384^3)$, medium ($448^3$) and high ($512^2$). Convergence performed on the amplitude and phase of the complex strain, $h_{22}$, extrapolated to null infinity. $\D t=0$ is the peak of the amplitude for the highest resolution. This figure indicates consistency with third order convergence.}
\label{fig:GW_strain_convergence}
\end{figure}

In Fig. \ref{fig:GW_strain_convergence} we show the error estimates in the quadrupole mode $h_{22}$ extrapolated to null infinity between low, medium and high resolutions and the estimates for the expected error assuming third and fourth order convergence. We decompose the complex strain into its amplitude and phase, $h_{\ell m} = h^+_{\ell m} - ih^\times_{\ell m} = h_{\ell m}^A e^{ih_{\ell m}^\f}$. We compute these expected errors using the continuum limit of the convergence factor of order $n$:
\begin{equation}
    Q_n = \frac{\br{\D x_{Low}}^n - \br{\D x_{Med}}^n}{\br{\D x_{Med}}^n - \br{\D x_{High}}^n}\,.
\end{equation}
This indicates the convergence order of $h_{22}$ is consistent with three. 
\begin{figure}[h]
\centering
\includegraphics[width=0.48\textwidth]{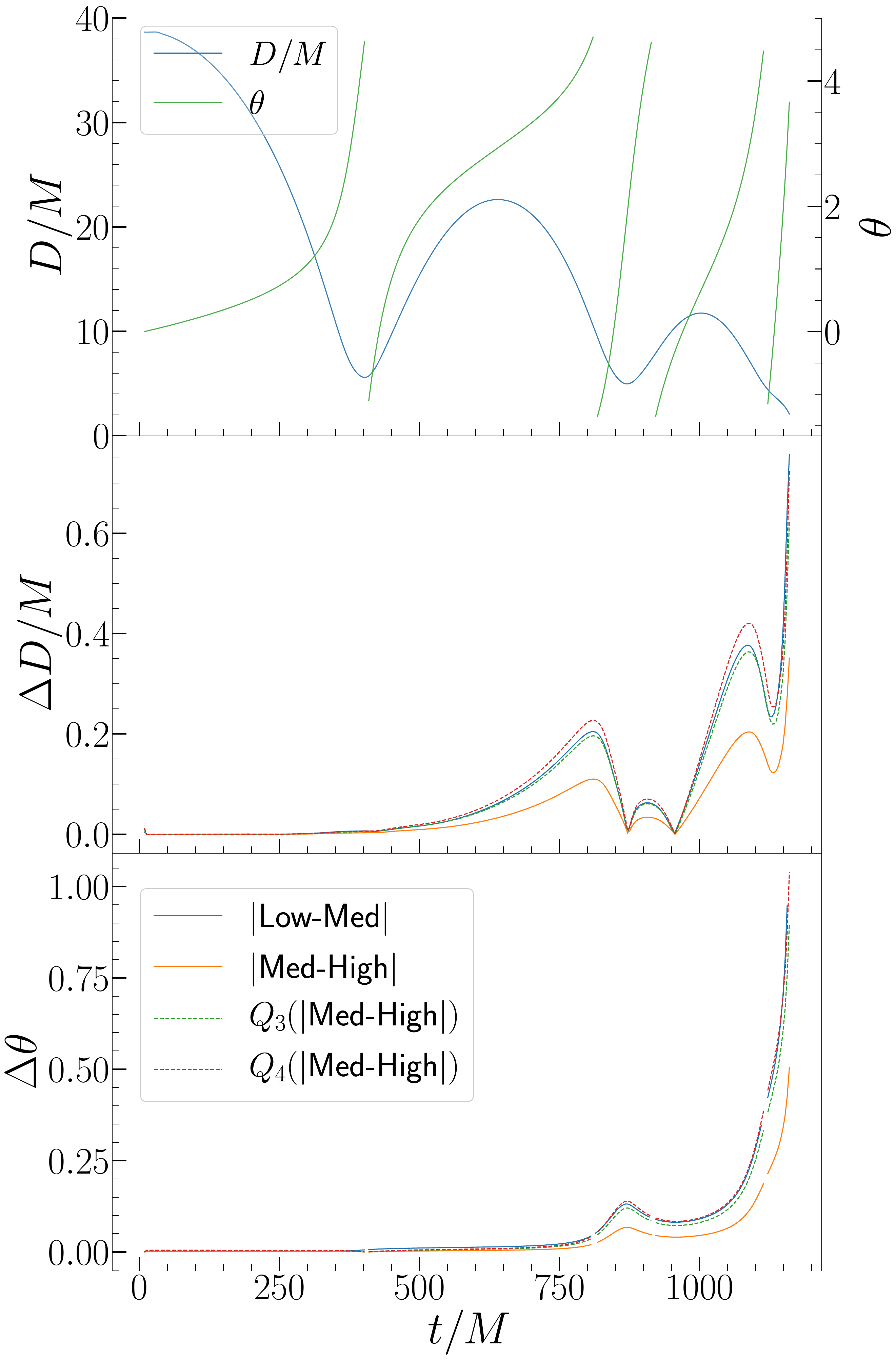}
\caption{\textit{Top panel:} Radial distance $D(t)$ and relative phase $\q(t)$ of the black holes' trajectories as functions of time for the $g_2=0.02$ Horndeski theory. Convergence tests for the radial distance $\D D(t)$ (\textit{middle panel}) and relative phase $\q(t)$ (\textit{bottom panel}). Both of these quantities exhibit between third and fourth order convergence.}
\label{fig:parametrize_2D_convergence}
\end{figure}

We also tested convergence of other variables; for instance, the trajectories of the two black holes, $x^i_{1}(t)$ and $x^i_2(t)$, shown in Fig. \ref{fig:AH_xy}, can be used to test convergence. We rewrite these trajectories in terms of the radial distance between the black holes,
\begin{equation}
    D(t)=|x^i_1(t) - x^i_2(t)|\,,
\end{equation} 
and the phase relative to the initial positions,
\begin{equation}
    \q(t)=\arccos\sbr{\frac{(x^i_1(t) - x^i_2(t))}{D(t)}\cdot\frac{(x^i_1(0) - x^i_2(0))}{D(0)}}\,.
\end{equation}
These quantities for the Horndeski theory are shown in the top panel of Fig. \ref{fig:parametrize_2D_convergence} for the same binary as in Fig.  \ref{fig:AH_xy}.
The convergence analysis of these quantities across the three resolutions is shown in the middle and bottom panels of Fig. \ref{fig:parametrize_2D_convergence}. These figures indicate that both quantities exhibit between third and fourth order convergence.
\begin{figure}[h]
\centering
\includegraphics[width=0.48\textwidth]{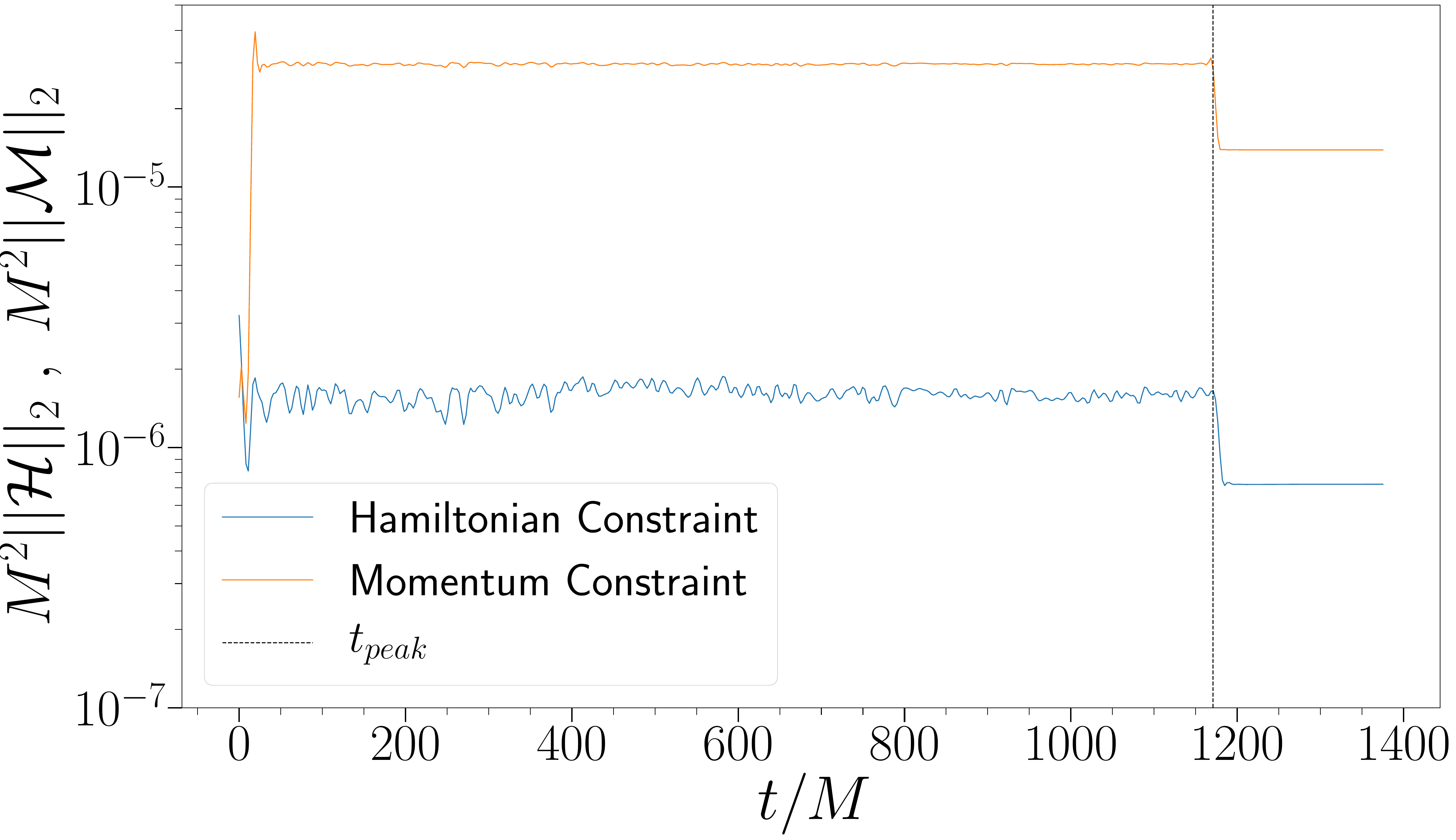}
\caption{$L^2$ norm of the Hamiltonian and the Euclidean norm of the momentum constraints for the medium resolution of the $g_2=0.02$ Horndeski binary.}
\label{fig:constraints}
\end{figure}

For completeness, in Fig. \ref{fig:constraints} we show the $L^2$ norms\footnote{For a given quantity $\mathcal{Q}$, we compute $L^2\mathcal{Q} = \sqrt{\frac{1}{V}\int_V|\mathcal{Q}^2|dV}$.} of the Hamiltonian and the Euclidean norm of the momentum constraints over the full computational domain. This figure shows the constraint violations remain stable at the level of $10^{-6}-10^{-5}M^{-2}$ respectively throughout the whole evolution, with a significant and sudden reduction at the merger. Considering the results of our convergence analysis, we conclude our simulations are stable and in the convergent regime.

\clearpage

%\nocite{*}

\bibliography{main}% Produces the bibliography via BibTeX.

\end{document}